\documentclass[journal]{IEEEtran}
\usepackage[T1]{fontenc}
\usepackage[latin9]{inputenc}
\usepackage{color}
\usepackage{array}
\usepackage{verbatim}
\usepackage{float}
\usepackage{booktabs}
\usepackage{multirow}
\usepackage{amsmath}
\usepackage{amsthm}
\usepackage{amssymb}
\usepackage{graphicx}
\usepackage[unicode=true,
 bookmarks=true,bookmarksnumbered=true,bookmarksopen=true,bookmarksopenlevel=1,
 breaklinks=false,pdfborder={0 0 0},pdfborderstyle={},backref=false,colorlinks=false]
 {hyperref}
\hypersetup{pdftitle={Your Title},
 pdfauthor={Your Name},
 pdfpagelayout=OneColumn, pdfnewwindow=true, pdfstartview=XYZ, plainpages=false}

\makeatletter

\providecommand{\tabularnewline}{\\}
\floatstyle{ruled}
\newfloat{algorithm}{tbp}{loa}
\providecommand{\algorithmname}{Algorithm}
\floatname{algorithm}{\protect\algorithmname}

\theoremstyle{plain}
\newtheorem{thm}{\protect\theoremname}
\theoremstyle{plain}
\newtheorem{lem}[thm]{\protect\lemmaname}

\usepackage[caption=false,font=normalsize]{subfig}
\DeclareMathOperator{\maximize}{maximize}
\DeclareMathOperator{\minimize}{minimize}
\DeclareMathOperator{\st}{subject~to}
\DeclareMathOperator{\diag}{diag}
\DeclareMathOperator{\tr}{Tr}
\DeclareMathOperator{\vect}{vec}

\usepackage{acronym}
\AtBeginDocument{\acrodef{AO}{alternating optimization}
\acrodef{ASP}{antenna separation product}
\acrodef{AWGN}{additive white Gaussian noise}
\acrodef{BEP}{bit error probability}
\acrodef{BER}{bit error rate}
\acrodef{BF-MIMO}[BF\mbox{-}MIMO]{beamforming MIMO}
\acrodef{BF}{beamforming}
\acrodef{bpcu}{bits per channel use}
\acrodef{CP}{cyclic prefix}
\acrodef{CSI}{channel state information}
\acrodef{CSIR}{channel state information at RX}
\acrodef{SSK}{space shift keying}
\acrodef{CSIT}{channel state information at TX}
\acrodef{DCMC}{discrete\mbox{-}input continuous\mbox{-}output memoryless channel}
\acrodef{DFT}{discrete Fourier transform}
\acrodef{DL-TR-GSM}{dual-layered transmit-receive \acl{GSM}}
\acrodef{DLT}{dual-layered transmission}
\acrodef{EGC}{equal gain combining}
\acrodef{EM}{electromagnetic}
\acrodef{FSPL}{free space path loss}
\acrodef{FFT}{fast Fourier transform}
\acrodef{FDE}{frequency domain equalization}
\acrodef{GRSM}{generalized \acl{RSM}}
\acrodef{GSM}{generalized \acl{SM}}
\acrodef{IFFT}{invserse fast Fourier transform}
\acrodef{ICI}{inter-channel interference}
\acrodef{iid}[i.i.d.]{independent and identically distributed}
\acrodef{IQ}{in\mbox{-}phase and quadrature}
\acrodef{ISI}{intersymbol interference}
\acrodef{ISI-free}[ISI\mbox{-}free]{intersymbol interference free}
\acrodef{LIS}{large intelligent surface}
\acrodef{LOS}{line\mbox{-}of\mbox{-}sight}
\acrodef{mmWave}{millimeter-wave}
\acrodef{MIMO}{multiple\mbox{-}input multiple\mbox{-}output}
\acrodef{MISO}{multiple\mbox{-}input single\mbox{-}output}
\acrodef{ML}{maximum likelihood}
\acrodef{MRC}{maximal ratio combining}
\acrodef{MMSE}{minimum mean square error}
\acrodef{MU-TR-GSM}{multiuser transmit-receive  \acl{GSM} }
\acrodef{NCSIT}{no channel state information at TX}
\acrodef{NLOS}{non\mbox{-}\acs{LOS}} 
\acrodef{NOMA}{non-orthogonal multiple access}
\acrodef{OFDM}{orthogonal frequency division multiplexing}
\acrodef{OFDMA}{orthogonal frequency division multiple access}
\acrodef{PA}{power amplifier}
\acrodef{PAE}{power added efficiency}
\acrodef{PAPR}{peak\mbox{-}to\mbox{-}average power ratio}
\acrodef{PDF}{probability density function}
\acrodef{PEP}{pairwise error probability}
\acrodefplural{PEP}{pairwise error probabilities}
\acrodef{PGM}{projected gradient method}
\acrodef{APGM}{accelerated projected gradient method}
\acrodef{PMP}{probability mass function}
\acrodef{PSM}{precoding-aided spatial modulation}
\acrodef{QSM}{quadrature spatial modulation}
\acrodef{RC}{reorganization computation}
\acrodef{RIS}{reconfigurable intelligent surface}
\acrodef{RSM}{receive spatial modulation}
\acrodef{RX}{receiver}
\acrodef{SEP}{symbol error probability}
\acrodef{SER}{symbol error rate}
\acrodef{SINR}{signal-to-interference-plus-noise ratio}
\acrodef{SISO}{single-input single-output}
\acrodef{SM}{spatial modulation}
\acrodef{SMX-MIMO}[SMX\mbox{-}MIMO]{spatial multiplexing MIMO}
\acrodef{SMX}{spatial multiplexing}
\acrodef{SNR}{signal-to-noise ratio}
\acrodef{SC}{single carrier}
\acrodef{SVD}{singular value decomposition}
\acrodef{SPST}{single pole single-throw}
\acrodef{SU}{secondary user}
\acrodef{TDE}{time domain equalization}
\acrodef{TX}{transmitter}
\acrodef{ULA}{uniform linear array}
\acrodef{URA}{uniform rectangular array}
\acrodef{VGA}{variable gain amplifier}
\acrodef{ZF}{zero-forcing}
\acrodef{ZMCG}{zero-mean complex Gaussian}

}
\usepackage{algorithm,algpseudocode}
\usepackage{setspace}

\@ifundefined{showcaptionsetup}{}{%
 \PassOptionsToPackage{caption=false}{subfig}}
\usepackage{subfig}
\makeatother

\providecommand{\lemmaname}{Lemma}
\providecommand{\theoremname}{Theorem}

\begin{document}
\title{\textcolor{black}{Achievable Rate Optimization for MIMO Systems with
Reconfigurable Intelligent Surfaces}}
\author{\textcolor{black}{Nemanja~Stefan~Perovi\'c, Le-Nam~Tran,~\IEEEmembership{\textcolor{black}{Senior Member,~IEEE,}}~Marco
Di Renzo,~\IEEEmembership{\textcolor{black}{Fellow,~IEEE,}} and~Mark~F.~Flanagan,~\IEEEmembership{\textcolor{black}{Senior~Member,~IEEE}}}\thanks{\textcolor{black}{The work of N. S. Perovi\'c and M. F. Flanagan
was funded by the Irish Research Council under grant number IRCLA/2017/209.
The work of L. N. Tran was supported in part by a Grant from Science
Foundation Ireland under Grant number 17/CDA/4786. M. Di Renzo's work
was supported in part by the European Commission through the H2020
ARIADNE project under grant agreement number 871464 and through the
H2020 RISE-6G project under grant agreement number 101017011.}}\textcolor{black}{}\thanks{\textcolor{black}{N. S. Perovi\'c, L. N. Tran, and M. F. Flanagan
are with School of Electrical and Electronic Engineering, University
College Dublin, Belfield, Dublin 4, Ireland (Email: nemanja.stefan.perovic@ucd.ie,
nam.tran@ucd.ie and mark.flanagan@ieee.org).}}\textcolor{black}{}\thanks{\textcolor{black}{M. Di Renzo is with Universit\'e Paris-Saclay,
CNRS, CentraleSup\'elec, Laboratoire des Signaux et Syst\`emes,
3 Rue Joliot-Curie, 91192 Gif-sur-Yvette, France (E-mail: marco.di-renzo@universite-paris-saclay.fr).}}}
\maketitle
\begin{abstract}
\textcolor{black}{Reconfigurable intelligent surfaces (RISs\acused{RIS})
represent a new technology that can shape the radio wave propagation
in wireless networks and offers a great variety of possible performance
and implementation gains. Motivated by this, we study the achievable
rate optimization for multi-stream \ac{MIMO} systems equipped with
an RIS, and formulate a joint optimization problem of the covariance
matrix of the transmitted signal and the RIS elements. To solve this
problem, we propose an iterative optimization algorithm that is based
on the \ac{PGM}. We derive the step size that guarantees the convergence
of the proposed algorithm and we define a backtracking line search
to improve its convergence rate. Furthermore, we introduce the total
\ac{FSPL} ratio of the indirect and direct links as a first-order
measure of the applicability of \acp{RIS} in the considered communication
system. Simulation results show that the proposed \ac{PGM} achieves
the same achievable rate as a state-of-the-art benchmark scheme, but
with a significantly lower computational complexity. In addition,
we demonstrate that the RIS application is particularly suitable to
increase the achievable rate in indoor environments, as even a small
number of RIS elements can provide a substantial achievable rate gain.\acresetall{}}
\end{abstract}

\begin{IEEEkeywords}
\textcolor{black}{Achievable rate, gradient projection, \ac{MIMO},
optimization, \ac{RIS}.\acresetall{}}
\end{IEEEkeywords}

\IEEEpeerreviewmaketitle{\textcolor{black}{}}

\section{\textcolor{black}{Introduction}}

\textcolor{black}{\bstctlcite{BSTcontrol}In recent years, there
has been a tremendous, almost exponential, increase in the demands
for higher data rates. The main driving forces that constantly increase
this demand are the increasing number of mobile devices and the appearance
of services that require high data rates (e.g., video streaming and
online gaming). Consequently, many technology solutions have been
proposed to address this ever-increasing demand, such as massive \ac{MIMO}
and \ac{mmWave} communications. In spite of providing potentially
significant achievable rate gains, these technologies generally incur
additional power and hardware costs, so that the total benefit of
their implementation has to be independently evaluated for each user
scenario. Broadly speaking, these technologies can be seen as novel
transmitter and receiver features that enable us to achieve higher
data rates. However, they do not have the capability of directly influencing
the propagation channel, the stochastic nature of which can sometimes
limit the efficiency of these proposed technology solutions.}

\textcolor{black}{A possible approach to overcome the aforementioned
issue lies in the use of the recently-developed \acp{RIS} \cite{di2019smart}.
The key component to realize the RIS function is a software-defined
surface that is reconfigurable in such a way as to adapt itself to
changes in the wireless environment. It consists of a large number
of small, low-cost, and passive elements, each of which can reflect
the incident signal with an adjustable phase shift, thereby modifying
the radio waves. Optimization of the wavefront of the reflected signals
enables us to shape how the radio waves interact with the surrounding
objects, and thus control their scattering and reflection characteristics
\cite{di2020reconfigurable,basar2019wireless,huang2020holographic}.
Hence, the introduction of \acp{RIS} fundamentally changes the
wave propagation in wireless communication systems and offers a wide
variety of possible implementation gains, thus potentially presenting
a new milestone in wireless communications.}

\textcolor{black}{In recent years, researchers have investigated many
important aspects of RIS-assisted wireless communication systems.
The problem of estimating the required \ac{CSI} was considered
in \cite{taha2019enabling} by embedding an active sensor in the RIS,
and in \cite{he2019cascaded} by estimation of a combined transmitter-RIS-receiver
channel. Other emerging body of work studies an accurate modeling
of the interactions (}considering\textcolor{black}{{} reflection, refraction},
diffraction\textcolor{black}{{} and polarization) of the incident wave
with the RIS, and elucidates the dependence of these interactions
on the size of the RIS elements, the distance between the adjacent
RIS elements, the angle of incidence and so on \cite{di2020smart,danufane2020path}.
All of these aspects are critical for the }practical\textcolor{black}{{}
implementation of RIS-aided wireless communication systems} to become
feasible\textcolor{black}{.}

\textcolor{black}{From a theoretical standpoint, the evaluation and
optimization of the achievable rate of an RIS-aided wireless communication
system is crucial. This problem is significantly more challenging
to solve than in the conventional case without an RIS, since in the
case without an RIS the channel capacity can be completely determined
in closed-form for deterministic (or fixed) channels. A variety of
different optimization methods for enhancing the achievable rate in
RIS-aided wireless communication systems have been proposed in the
literature, which attempt to find a near-optimal solution with a reasonable
computational complexity and run time. The vast majority of these
methods are particularly tailored for downlink communication with
single-antenna receive devices. In \cite{wu2019intelligent}, the
authors introduced an optimization method that increases the receive
\ac{SNR} and consequently enhances the achievable rate in \ac{MISO}
systems. The proposed solution is based on the \ac{AO} method,
which adjusts the transmit beamformer and the RIS element phase shifts
in an alternating fashion. The AO technique has also been successfully
utilized to increase the data rate for secure communications in environments
with multiple \acp{RIS} and single-antenna users \cite{yu2019robust}.
In contrast to \ac{AO}, the spectral efficiency optimization for
a single-user MISO system in \cite{Yu_2019} was performed by jointly
adjusting the transmit beamformer and the RIS element phase shifts.
In \cite{kammoun2020asymptotic}, the authors employed a gradient-based
algorithm to enhance the receive \ac{SINR}, and hence the achievable
rate, for single-antenna users that do not have a direct link with
the base station. The achievable rate optimization for multi-user
downlink communications is specifically considered for \ac{mmWave}
sparsely scattered channels in \cite{wang2019intelligent}. An algorithm
for energy efficiency optimization in a multi-user downlink communication
system was presented in \cite{huang2019reconfigurable}. The sum-rate
optimization for multi-user downlink communications using a deep reinforcement
learning based algorithm was introduced in \cite{huang2020reconfigurable}.}

\textcolor{black}{In contrast to the previous papers, the achievable
rate optimization in \cite{zhao2020exploiting} is realized by jointly
controlling the phase and the amplitude adjustment of each RIS element.
Furthermore, in \cite{abeywickrama2020intelligent} the authors developed
a practical phase shift model that captures the phase-dependent amplitude
variation in the RIS element-wise reflection coefficient and utilized
it to enhance the achievable rate. The achievable rate optimization
for an RIS with discrete phase shifts in multi-user downlink communications
was considered in \cite{di2020hybrid}. A system for serving paired
power-domain \ac{NOMA} users by designing the RIS phase shifts
was introduced in \cite{hou2020reconfigurable}. In \cite{yang2019irs},
the authors studied the joint optimization of the RIS reflection coefficients
and the \ac{OFDMA} time-frequency resource block, as well as power
allocations, to maximize the users' common (minimum) rate. Energy-efficiency
optimization for multi-user uplink MIMO communications was presented
in \cite{xiong2020reconfigurable}, where the users' covariance matrices
and the RIS phase shifts are optimized in an alternating fashion based
on partial knowledge of the \ac{CSI}.}

\textcolor{black}{Although RIS-aided communication systems with single-antenna
devices are well-studied in the literature, there is only a limited
number of papers that consider the }design and\textcolor{black}{{} analysis
of an RIS-aided }\textcolor{black}{\emph{MIMO}}\textcolor{black}{{}
communication system. In particular, the achievable rate optimization
in those systems remains relatively unknown. It was demonstrated in
\cite{ozdogan2020using} how an RIS can be implemented and optimized
to increase the rank of the channel matrix, leading to substantial
achievable rate gains in }\textcolor{black}{\emph{multi-stream}}\textcolor{black}{{}
\ac{MIMO} communications. The method proposed in \cite{ozdogan2020using}
was specifically designed for pure \ac{LOS} channels, neglecting
the presence of any \ac{NLOS} component. Optimization of the achievable
rate for a }\textcolor{black}{\emph{single-stream}}\textcolor{black}{{}
MIMO system in an indoor \ac{mmWave} environment with a blocked
direct link was analyzed in~\cite{perovic2019channel}. Since in
indoor \ac{mmWave} communications \ac{NLOS} channel components
are usually significantly weaker than the \ac{LOS} component, all
communication links were also modeled as pure LOS links. Although
the proposed optimization schemes in \cite{perovic2019channel} provide
the near-optimal achievable rate, they require very low computational
and hardware complexity. In \cite{zhang2019capacity}, the authors
utilized the \ac{AO} method to enhance the achievable rate of an
RIS-aided }\textcolor{black}{\emph{multi-stream}}\textcolor{black}{{}
MIMO communication system. Although this optimization method is simple
to implement, it can require many iterations to converge, especially
when the number of RIS elements is very large, which corresponds precisely
to the case where the RIS is the most useful. }

\textcolor{black}{Against this background, the contributions of this
paper are listed as follows:}
\begin{itemize}
\item \textcolor{black}{To maximize the achievable rate of a multi-stream
MIMO system equipped with an RIS, we formulate a joint optimization
problem of the covariance matrix of the transmitted signal and the
RIS elements (i.e., phase shifts). We then propose an iterative \ac{PGM}
to solve this nonconvex problem, for which we present exact gradient
and projection expressions in closed form. The proposed method is
provably convergent to a critical point of the considered problem,
which is desirable for a nonconvex program.}
\item \textcolor{black}{We derive a Lipschitz constant for the proposed
PGM, which is then used to determine an appropriate step size which
guarantees its convergence. Also, to improve the rate of convergence
of the proposed algorithm, we propose a data scaling step and employ
a backtracking line search, which increases the convergence rate significantly,
and more importantly, outperforms the existing AO approach in terms
of convergence rate.}
\item As a tool\textcolor{black}{{} to estimate the applicability of an RIS,
we introduce the concept of the }\textcolor{black}{\emph{total free
space path loss}}\textcolor{black}{{} (total {FSPL}\acused{FSPL}).
Since the computation of the total \acs{FSPL} of the indirect link
is an intractable problem in a \ac{MIMO} system, we instead derive
the total \acp{FSPL} for a \ac{SISO} system. We }then\textcolor{black}{{}
show that the ratio of the total FSPL of the indirect and direct links
can be used as an accurate first-order measure of the applicability
of an RIS.}
\item \textcolor{black}{We show through simulations that the proposed \ac{PGM}
provides the same achievable rate as the AO, but with a significantly
lower number of iterations. This is particularly visible in the case
where the direct link is blocked, as in this case the PGM needs just
a few iterations to reach the convergent achievable rate. As a side
product, we demonstrate that the total \ac{FSPL} of the indirect
link is primarily determined by the RIS position, while the total
length of the indirect link is of relatively minor importance. Also,
we show that scaling the number of RIS elements with the operational
frequency can compensate the FSPL increase and ensure communication
via the indirect link at all frequencies. Furthermore, we study the
application of an RIS in an indoor environment and show that a small
number of RIS elements is sufficient to enable the indirect link to
have a higher achievable rate than the direct link. Last but not least,
we demonstrate that the proposed \ac{PGM} has a significantly lower
computational complexity compared to the AO method.}
\end{itemize}
The rest of this paper is organized as follows. In Section \ref{sec:System-Model},
we introduce the system model and formulate the optimization problem
to maximize the achievable rate of a MIMO system equipped with an
RIS. In Section \ref{sec:Solution-Approach-PGM}, we propose and derive
the \ac{PGM} algorithm to solve the previous optimization problem.
The convergence and the complexity analysis of the proposed optimization
algorithm are presented in Section \ref{sec:Convergence-and-Complexity}.
The applicability of an RIS in the considered communication system
is discussed in Section \ref{sec:Total-FSPL}. In Section~\ref{sec:Simulation-Results},
we illustrate simulation results of the achievable rate for the proposed
\ac{PGM} algorithm, and use these to illustrate its advantages.
Finally, Section \ref{sec:Conclusion} concludes this paper.

\paragraph*{\textcolor{black}{Notation}}

\textcolor{black}{Bold lower and upper case letters represent vectors
and matrices, respectively. $\mathbb{C}^{a\times b}$ denotes the
space of complex matrices of dimensions $a\times b$. $\ensuremath{(\cdot)^{T}}$,
$\ensuremath{(\cdot)^{\ast}}$ and $\ensuremath{(\cdot)^{H}}$ represent
transpose, complex conjugate and Hermitian transpose, respectively.
$\ln(x)$ denotes the natural logarithm of $x$. $\lambda_{\max}(\mathbf{X})$
denotes the largest singular value of matrix $\mathbf{X}$. To simplify
the notation we denote by $||\cdot||$ the Euclidean norm if the argument
is a vector and the Frobenius norm if the argument is a matrix. $\mathrm{diag}\left(\mathbf{x}\right)$
denotes the square diagonal matrix which has the elements of $\mathbf{x}$
on the main diagonal. $\left|x\right|$ is the absolute value of $x$
and $(x)_{+}$ denotes $\max(0,x)$. $\mathrm{arg}\{x\}$ denotes
the argument of $x.$ The $l$-th entry of vector $\mathbf{x}$ is
denoted by $x_{l}$. $\tr(\mathbf{X})$ is the trace of matrix $\mathbf{X}$,
and $\mathbb{E}\{\cdot\}$ stands for the expectation operator. $\det(\mathbf{X})$
is the determinant of $\mathbf{X}$. The notation $\mathbf{A}\succeq(\succ)\mathbf{B}$
means that $\mathbf{A}-\mathbf{B}$ is positive semidefinite (definite).
$\nabla_{\mathbf{X}}f(\cdot)$ is the gradient of $f$ with respect
to $\mathbf{X}^{*}\in\mathbb{C}^{m\times n}$, which also lies in
$\mathbb{C}^{m\times n}$. $\vect_{d}(\mathbf{X})$ denotes the vector
comprised of the diagonal elements of $\mathbf{X}$. $\vect(\mathbf{X})$
denotes the vectorization operator which stacks the columns of $\mathbf{X}$
to create a single long column vector. $A(i,k)$ denotes the $k$-th
element of the $i$-th row of matrix $\mathbf{A}$.}

\section{\textcolor{black}{System Model and Problem Formulation\label{sec:System-Model}}}

\subsection{\textcolor{black}{System Model}}

\textcolor{black}{}
\begin{figure}[t]
\centering{}\textcolor{black}{\includegraphics[scale=0.9]{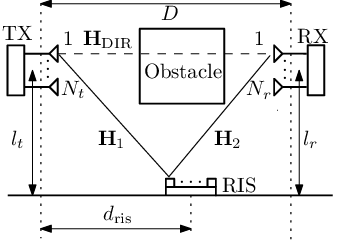}\caption{Aerial view of the considered communication system. \label{fig:System_model}}
}
\end{figure}
\textcolor{black}{We consider a wireless communication system with
$N_{t}$ transmit and $N_{r}$ receive antennas, whose aerial view
is depicted in Fig. \ref{fig:System_model}. Both the transmit and
receive antennas are placed in \acp{ULA} on vertical walls that are
parallel to each other. The distance between these walls is denoted
by $D$. For simplicity, both antenna arrays are parallel to the ground
and are assumed to be at the same height. The inter-antenna separations
of these arrays are denoted by $s_{t}$ and $s_{r}$, respectively.
The direct link is attenuated by an obstacle (e.g., a building) which
is situated between the two antenna arrays, and for this reason, a
rectangular RIS of size $a\times b$ is utilized to improve the system
performance. The RIS is installed on a vertical wall that is perpendicular
to the antenna arrays  and its center is at the same height as the
transmit and the receive antenna arrays}\footnote{\textcolor{black}{For ease of exposition, we assume that the RIS,
the transmit and the receive antenna arrays are at the same height.
It can be shown by simulations that introducing different heights
has a negligible influence on the achievable rate.}}\textcolor{black}{. It consists of reflection elements placed in an
\ac{URA} with $N_{a}$ and $N_{b}$ elements per dimension respectively
(the total number of reflection elements of the RIS then being $N_{\mathrm{ris}}=N_{a}N_{b}$).
All RIS elements are of size $\frac{\lambda}{2}\times\frac{\lambda}{2}$,
where $\lambda$ denotes the wavelength of operation. The separation
between the centers of adjacent RIS elements in both dimensions is
$s_{\mathrm{ris}}=\frac{\lambda}{2}$. The distance between the midpoint
of the RIS and the plane containing the transmit antenna array is
$d_{\mathrm{ris}}$. The distance between the midpoint of the transmit
antenna array and the plane containing the RIS is $l_{t}$, and the
distance between the midpoint of the receive antenna array and the
plane containing the RIS is $l_{r}$. We assume that the RIS elements
are ideal and that each of them can independently influence the phase
and the reflection angle of the impinging wave.}

\textcolor{black}{The signal vector at the receive antenna array is
given by
\begin{equation}
\mathbf{y}=\mathbf{Hx}+\mathbf{n},\label{eq:ss_equ-1}
\end{equation}
where $\mathbf{H}\in\mathbb{C}^{N_{r}\times N_{t}}$ is the channel
matrix, $\mathbf{x}\in\mathbb{C}^{N_{t}\times1}$ is the transmit
signal vector and $\mathbf{n}\in\mathbb{C}^{N_{r}\times1}$ is the
noise vector which is distributed according to $\mathcal{CN}(\mathbf{0},N_{0}\mathbf{I})$.
We assume that the total average transmit power has a maximum value
of $P_{t}$, i.e., $\mathbb{E}\{\mathbf{x}^{H}\mathbf{x}\}\le P_{t}$.
Let $\mathbf{\ensuremath{Q}}\succeq\mathbf{0}$ be the covariance
matrix of the transmitted signal, i.e., $\mathbf{\ensuremath{Q}}=\mathbb{E}\{\mathbf{x}\mathbf{x}^{H}\}$,
then the transmit power constraint can be equivalently written }as
\begin{equation}
\tr(\mathbf{\ensuremath{Q}})\le P_{t}.\label{eq:pow_con}
\end{equation}

\subsection{Channel Model}

\textcolor{black}{Since an RIS is present in this system, the channel
matrix can be expressed as
\[
\mathbf{H}=\mathbf{H}_{\mathrm{DIR}}+\mathbf{H}_{\mathrm{INDIR}},
\]
where $\mathbf{H}_{\mathrm{DIR}}\in\mathbb{C}^{N_{r}\times N_{t}}$
represents the }\textcolor{black}{\emph{direct }}\textcolor{black}{link
between the transmitter and the receiver, and $\mathbf{H}_{\mathrm{INDIR}}\in\mathbb{C}^{N_{r}\times N_{t}}$
represents the }\textcolor{black}{\emph{indirect }}\textcolor{black}{link
between the transmitter and the receiver (i.e., via the RIS). Adopting
the Rician fading channel model,  the direct link channel matrix
is given by
\begin{equation}
\mathbf{H}_{\mathrm{DIR}}=\frac{\sqrt{\beta_{\mathrm{DIR}}^{-1}}}{\sqrt{K+1}}(\sqrt{K}\mathbf{H}_{\mathrm{D,}\mathrm{LOS}}+\mathbf{H}_{\mathrm{D},\mathrm{NLOS}}),
\end{equation}
where $H_{\mathrm{D,}\mathrm{LOS}}(r,t)=e^{-j2\pi d_{r,t}/\lambda}$
and $d_{r,t}$ is the distance between the $t$-th transmit and the
$r$-th receive antenna. The elements of $\mathbf{H}_{\mathrm{D,\mathrm{NLOS}}}$
are \ac{iid} according to $\mathcal{CN}(0,1)$. The \ac{FSPL} for
the direct link is given by $\beta_{\mathrm{DIR}}=(4\pi/\lambda)^{2}d_{0}^{\alpha_{\mathrm{DIR}}}$
\cite{rappaport2017overview}, where $d_{0}=\sqrt{D^{2}+(l_{t}-l_{r})^{2}}$
is the distance between the transmit array midpoint and the receive
array midpoint. The path loss exponent of the direct link, whose value
is influenced by the obstacle present, is denoted by $\alpha_{\mathrm{DIR}}$.
The Rician factor $K$ is chosen from the interval $[0,+\infty)$.}

\textcolor{black}{We assume that the far-field model is valid for
signal transmission via the RIS (i.e., for the indirect link), and
thus $\mathbf{H}_{\mathrm{INDIR}}$ can be written as
\begin{equation}
\mathbf{H}_{\mathrm{INDIR}}=\sqrt{\beta_{\mathrm{INDIR}}^{-1}}\mathbf{H}_{2}\mathbf{F}(\boldsymbol{\theta})\mathbf{H}_{1},\label{eq:INDIR_prod}
\end{equation}
where $\mathbf{H}_{1}\in\mathbb{C}^{N_{\mathrm{ris}}\times N_{t}}$
represents the channel between the transmitter and the RIS, $\mathbf{H}_{2}\in\mathbb{C}^{N_{r}\times N_{\mathrm{ris}}}$
represents the channel between the RIS and the receiver, and $\beta_{\mathrm{INDIR}}^{-1}$
represents the overall \ac{FSPL} for the indirect link. Signal reflection
from the RIS is modeled by the matrix $\mathbf{F}(\boldsymbol{\theta})=\mathrm{diag}(\boldsymbol{\theta})\in\mathbb{C}^{N_{\mathrm{ris}}\times N_{\mathrm{ris}}}$,
where $\boldsymbol{\theta}=[\theta_{1},\theta_{2},\ldots,\theta_{N_{\mathrm{ris}}}]^{T}\in\mathbb{C}^{N_{\mathrm{ris}}\times1}$.
In this paper, similar to related works \cite{wu2019intelligent,zhang2019capacity},
we assume that the signal reflection from any RIS element is ideal,
i.e., without any power loss. }In other words, we may write $\theta_{l}=e^{j\phi_{l}}$
for $l=1,2,\ldots,N_{\mathrm{ris}}$, where $\phi_{l}$ is the phase
shift induced by the $l$-th RIS element. Equivalently, we may write
\begin{equation}
\left|\theta_{l}\right|=1,\quad l=1,2,\ldots,N_{\mathrm{ris}}.\label{eq:RIS_elem_cons}
\end{equation}

Utilizing the Rician fading channel model, the channel between the
transmitter and the RIS $\mathbf{H}_{1}$ is given by\textcolor{black}{
\begin{equation}
\mathbf{H}_{1}=\frac{1}{\sqrt{K+1}}(\sqrt{K}\mathbf{H}_{\mathrm{1,LOS}}+\mathbf{H}_{\mathrm{1,NLOS}}),\label{eq:H1_expr}
\end{equation}
where $H_{1,\mathrm{LOS}}(l,t)=e^{-j2\pi d_{l,t}/\lambda}$ and $d_{l,t}$
is the distance between the $t$-th transmit antenna and the $l$-th
RIS element. The elements of $\mathbf{H}_{1,\mathrm{NLOS}}$ are \ac{iid}
according to $\mathcal{CN}(0,1)$. It is worth noting that the channel
matrix expression \eqref{eq:H1_expr} does not contain any \ac{FSPL}
term.}

In a similar way, $\mathbf{H}_{2}$ can be expressed as
\begin{equation}
\mathbf{H}_{2}=\sqrt{\frac{1}{K+1}}(\sqrt{K}\mathbf{H}_{\mathrm{2,LOS}}+\mathbf{H}_{\mathrm{2,NLOS}})
\end{equation}
\textcolor{black}{where $\mathbf{H}_{2,\mathrm{LOS}}(r,l)=e^{-j2\pi d_{r,l}/\lambda}$
and $d_{r,l}$ is the distance between the $l$-th RIS element and
the $r$-th receive antenna. The \ac{FSPL} for the indirect link
can be computed according to \cite[Eqn. (18.13.6)]{danufane2020path,ellingson2019path,orfanidis2002electromagnetic}
as
\begin{equation}
\beta_{\mathrm{INDIR}}^{-1}=\frac{\lambda^{4}}{256\pi^{2}}\frac{(\cos\gamma_{1}+\cos\gamma_{2})^{2}}{d_{1}^{2}d_{2}^{2}},\label{eq:FSPL_indir}
\end{equation}
where $d_{1}=\sqrt{d_{\mathrm{ris}}^{2}+l_{t}^{2}}$ is the distance
between the transmit array midpoint and the RIS center, and $d_{2}=\sqrt{(D-d_{\mathrm{ris}})^{2}+l_{r}^{2}}$
is the distance between the RIS center and the receive array midpoint.
Also, $\gamma_{1}$ is the angle between the incident wave direction
from the transmit array midpoint to the RIS center and the vector
normal to the RIS, and $\gamma_{2}$ is the angle between the vector
normal to the RIS and the reflected wave direction from the RIS center
to the receive array midpoint. Therefore, we have $\cos\gamma_{1}=l_{t}/d_{1}$
and $\cos\gamma_{2}=l_{r}/d_{2}$, which finally gives }
\begin{equation}
\beta_{\mathrm{INDIR}}^{-1}=\frac{\lambda^{4}}{256\pi^{2}}\frac{(l_{t}/d_{1}+l_{r}/d_{2})^{2}}{d_{1}^{2}d_{2}^{2}}.
\end{equation}

\subsection{Problem Formulation}

\textcolor{black}{In this paper, we are interested in maximizing the
achievable rate}\footnote{\textcolor{black}{Note that this achievable rate does not correspond
to the channel \emph{capacity}, as we do not consider the possibility
of encoding the transmitted data into the phase shift values of the
RIS. If such encoding is performed, the capacity of the RIS-aided
MIMO system may be achieved \cite{karasik2019beyond}.}}\textcolor{black}{{} of the considered RIS-assisted wireless communication
system. It is well known that for a MIMO channel, Gaussian signaling
provides the maximum achievable rate, and that for a given input covariance
matrix $\mathbf{Q}$, when $\mathbf{H}$ is known perfectly at both
transmitter and receiver, the following rate is achievable:
\begin{equation}
R=\log_{2}\det\Bigl(\mathbf{I}+\frac{1}{N_{0}}\mathbf{H}\mathbf{Q}\mathbf{H}^{H}\Bigr)(\textrm{bit/s/Hz}).\label{eq:Cap_def}
\end{equation}
We note that the channel matrix $\mathbf{H}$ also depends on $\boldsymbol{\theta}$.
Thus, for the total power $P_{t}$, the problem of the achievable
rate optimization for the considered system can be mathematically
stated as:\begin{subequations}\label{eq:capacityprob}
\begin{align}
\underset{\boldsymbol{\theta},\mathbf{Q}}{\maximize} & \ f(\boldsymbol{\theta},\mathbf{Q})=\ln\det\Bigl(\mathbf{I}+\mathbf{Z}(\boldsymbol{\theta})\mathbf{Q}\mathbf{Z}^{H}(\boldsymbol{\theta})\Bigr)\label{eq:opt_problem}\\
\st & \ \tr(\mathbf{Q})\le P_{t};\mathbf{Q}\succeq\mathbf{0};\\
 & \ \bigl|\theta_{l}\bigr|=1,l=1,2,\ldots,N_{\mathrm{ris}}.
\end{align}
\end{subequations}where
\begin{align}
\mathbf{Z}(\boldsymbol{\theta}) & =\bar{\mathbf{H}}_{\mathrm{DIR}}+\mathbf{H}_{2}\mathbf{F}(\boldsymbol{\theta})\mathbf{\bar{\mathbf{H}}}_{1}\\
\bar{\mathbf{H}}_{\mathrm{DIR}} & =\mathbf{H}_{\mathrm{DIR}}/\sqrt{N_{0}}\label{eq:Hdir}\\
\bar{\mathbf{H}}_{\mathrm{1}} & =\mathbf{H}_{\mathrm{1}}\sqrt{\beta_{\mathrm{INDIR}}^{-1}/N_{0}}.
\end{align}
}

\section{Solution Approach via Projected Gradient Method\label{sec:Solution-Approach-PGM}}

\textcolor{black}{In contrast to the conventional MIMO channel where
the water-filling algorithm can be used to efficiently find the maximum
achievable rate, problem \eqref{eq:capacityprob} is nonconvex and
thus difficult to solve. Further, we note that the objective is neither
convex nor concave in the involved variables.}

\textcolor{black}{Previously proposed methods for rate optimization
in RIS communication systems were primarily based on the }\textcolor{black}{\emph{alternating
optimization }}\textcolor{black}{(AO)\acused{AO} technique \cite{wu2019intelligent,zhang2019capacity}.
The main idea of this method is that the RIS phase shifts and the
covariance matrix are optimized in an alternating fashion, each independently
of the other. This method is motivated by the fact that the optimization
over one variable can be performed efficiently (i.e., in closed form)
while others are kept fixed. Although the AO method is easy to implement,
it may require many iterations to converge, especially when the number
of RIS elements is very large (which corresponds to the case in which
the RIS is the most useful). In other words, the simplicity of an
iteration in the AO method does not necessarily translate into low
actual run time.}

\textcolor{black}{Motivated by the above discussion, we propose an
optimization method to solve \eqref{eq:capacityprob}, based on the
\ac{PGM} presented in \cite{Li:2015:APG}. Our proposed method is
motivated by the fact that the projection onto the feasible set (albeit
nonconvex with respect to $\boldsymbol{\theta}$) can be performed
efficiently.}

\subsection{\textcolor{black}{Description of Proposed Algorithm}}

\textcolor{black}{To describe the proposed algorithm, we define the
following two sets:
\begin{equation}
\Theta=\{\boldsymbol{\theta}\in\mathbb{C}^{N_{\mathrm{ris}}\times1}:\bigl|\theta_{l}\bigr|=1,l=1,2,\ldots,N_{\mathrm{ris}}\}
\end{equation}
\begin{equation}
\mathcal{Q}=\{\mathbf{Q}\in\mathbb{C}^{N_{t}\times N_{t}}:\tr(\mathbf{Q})\le P_{t};\mathbf{Q}\succeq\mathbf{0}\}
\end{equation}
It is clear that the feasible set of \eqref{eq:capacityprob} is the
Cartesian product of $\Theta$ and $\mathcal{Q}$. We denote by $P_{\mathcal{U}}(\mathbf{u})$
the Euclidean projection from a point $\mathbf{u}$ onto a set $\mathcal{U}$,
i.e., $P_{\mathcal{U}}(\mathbf{u})=\underset{\mathbf{x}}{\arg\min}\{||\mathbf{x}-\mathbf{u}||:\mathbf{x}\in\mathcal{U}\}$.}

\textcolor{black}{The proposed algorithm is outlined in Algorithm
\ref{alg:GPA} and follows the projected gradient method in order
to solve \eqref{eq:capacityprob}. The main idea behind Algorithm
\ref{alg:GPA} is as follows. Starting from an arbitrary point $(\boldsymbol{\theta}_{0},\mathbf{Q}_{0})$,
we move in each iteration in the direction of the gradient of $f(\boldsymbol{\theta},\mathbf{Q})$.
The size of this move is determined by the step size $\mu>0$ (see
Section \ref{sec:Convergence-and-Complexity} for details regarding
the choice of an appropriate step size). As a result of this step,
the resulting updated point may lie outside of the feasible set. Therefore,
before the next iteration, we project the newly computed points $\boldsymbol{\theta}$
and $\mathbf{Q}$ onto $\Theta$ and $\mathcal{Q}$, respectively.
As shall be seen shortly, the projection onto $\Theta$ or $\mathcal{Q}$
can be determined in closed form. Another important remark concerning
Algorithm \ref{alg:GPA} is also in order. Since \eqref{eq:capacityprob}
involves complex variables, we adopt the complex-valued gradient defined
in \cite[ Eq. (4.37)]{Are2011}. In particular, it is proved that
the directions where $f(\boldsymbol{\theta},\mathbf{Q})$ has maximum
rate of change with respect to $\boldsymbol{\theta}$ and $\mathbf{Q}$
are $\nabla_{\boldsymbol{\theta}}f(\boldsymbol{\theta},\mathbf{Q})$
and $\nabla_{\mathbf{Q}}f(\boldsymbol{\theta},\mathbf{Q})$, respectively
\mbox{\cite[Theorem 3.4]{Are2011}}.}

\textcolor{black}{In our method, all optimization variables are updated
simultaneously in each iteration. This is in sharp contrast to the
AO method, in which each iteration only updates a single variable.
As a result, the proposed method converges much faster than the AO
method as we demonstrate via extensive numerical results in Section
\ref{sec:Simulation-Results}. }

\subsection{\textcolor{black}{Complex-valued Gradient of $f(\boldsymbol{\theta},\mathbf{Q})$}}

\textcolor{black}{Let $\mathbf{K}(\boldsymbol{\theta},\mathbf{Q})=(\mathbf{I}+\mathbf{Z}(\boldsymbol{\theta})\mathbf{Q}\mathbf{Z}^{H}(\boldsymbol{\theta}))^{-1}$.
Then we have the following result.}
\begin{lem}
\textcolor{black}{The gradient of $f(\boldsymbol{\theta},\mathbf{Q})$
with respect to $\boldsymbol{\theta}^{\ast}$ and $\mathbf{Q}^{\ast}$
is given by}\begin{subequations}
\begin{align}
\nabla_{\boldsymbol{\theta}}f(\boldsymbol{\theta},\mathbf{Q}) & =\vect_{d}\left(\mathbf{H}_{2}^{H}\mathbf{K}(\boldsymbol{\theta},\mathbf{Q})\mathbf{Z}(\boldsymbol{\theta})\mathbf{Q}\bar{\mathbf{H}}_{1}^{H}\right)\label{eq:grad_theta}\\
\nabla_{\mathbf{Q}}f(\boldsymbol{\theta},\mathbf{Q}) & =\mathbf{Z}^{H}(\boldsymbol{\theta})\mathbf{K}(\boldsymbol{\theta},\mathbf{Q})\mathbf{Z}(\boldsymbol{\theta}).\label{eq:grad_Q}
\end{align}
\end{subequations}
\end{lem}
\begin{IEEEproof}
See Appendix \ref{sec:grad}.
\end{IEEEproof}

\subsection{\textcolor{black}{Projection onto $\Theta$ and $\mathcal{Q}$}}

\textcolor{black}{We now show that the projection operations in Algorithm~\ref{alg:GPA}
can be carried out very efficiently, and thus Algorithm \ref{alg:GPA}
indeed requires low complexity to implement. Note that the constraint
$\bigl|\theta_{l}\bigr|=1$ means that $\theta_{l}$ should lie on
the unit circle in the complex plane. Thus, it is straightforward
to see that, for a given point $\mathbf{u}\in\mathbb{C}^{N_{\mathrm{ris}}\times1}$,
$P_{\Theta}(\mathbf{u})$ is the vector $\bar{\mathbf{u}}$ where
\begin{equation}
\bar{u}_{l}=\begin{cases}
\frac{u_{l}}{|u_{l}|} & u_{l}\neq0\\
e^{j\phi},\phi\in[0,2\pi] & u_{l}=0
\end{cases},l=1,\dots,N_{\mathrm{ris}}.\label{eq:projectthetha}
\end{equation}
Note that $\bar{u}_{l}$ can be any point on the unit circle if $u_{l}=0$,
and thus the projection onto $\Theta$ is not unique. Despite this
issue, we are still able to prove the convergence of \mbox{Algorithm
\ref{alg:GPA}}, which is shown in the next section. Next we turn
our attention to the projection onto $\mathcal{Q}$, which general
problem has already been studied previously (e.g., in \cite{pham2018revisiting}).
For a given $\mathbf{Y}\succeq\mathbf{0}$, the projection of $\mathbf{Y}$
onto $\mathcal{Q}$ is the solution of the following problem:\begin{subequations}\label{eq:projectQ}
\begin{align}
\underset{\mathbf{Q}}{\minimize} & \quad\left\Vert \mathbf{Q}-\mathbf{Y}\right\Vert ^{2}\\
\st & \quad\tr(\mathbf{Q})\le P_{t};\mathbf{Q}\succeq\mathbf{0}
\end{align}
\end{subequations}Let $\mathbf{Y}=\mathbf{U}\boldsymbol{\Sigma}\mathbf{U}^{H}$
be the eigenvalue decomposition of $\mathbf{Y}$, where $\boldsymbol{\Sigma}=\diag(\sigma_{1},\dots,\sigma_{N_{t}})$.
Now, we can write $\mathbf{Q}=\mathbf{UD}\mathbf{U}^{H}$ for some
$\mathbf{D}\succeq\mathbf{0}$ and $\tr(\mathbf{Q})=\tr(\mathbf{D})$.
Then, we obtain $\left\Vert \mathbf{Q}-\mathbf{Y}\right\Vert ^{2}=\left\Vert \boldsymbol{\Sigma}-\mathbf{D}\right\Vert ^{2}$.
Thus, $\mathbf{D}$ must be diagonal to be an optimal solution, i.e.,
$\mathbf{D}=\diag(d_{1},\dots,d_{N_{t}})$. Therefore, \eqref{eq:projectQ}
is equivalent to the following program:\begin{subequations}
\begin{align}
\underset{\{d_{i}\}}{\minimize} & \quad\sum\nolimits _{i=1}^{N_{t}}(d_{i}-\sigma_{i})^{2}\\
\st & \quad\sum\nolimits _{i=1}^{N_{t}}d_{i}\le P_{t};d_{i}\ge0
\end{align}
\end{subequations}The solution to the above problem is achieved by
the water-filling algorithm and is given as
\begin{equation}
d_{i}=(\sigma_{i}-\gamma)_{+},\quad i=1,\dots,N_{t},\label{eq:WF}
\end{equation}
where $\gamma\ge0$ is the water level.}

\subsection{\textcolor{black}{Improved Convergence Rate by Data Scaling\label{subsec:Imp-Conv}}}

\begin{algorithm}[t]
\caption{Proposed projected gradient method (PGM). \label{alg:GPA}}

\begin{algorithmic}[1] 
\State{$\mathbf{Input:} \; \mathbf{\theta}_{0},\mathbf{Q}_{0},\mu>0.$}
\For {$n=1,2,\ldots$}
\State{$\mathbf{\theta}_{n+1}=P_{\Theta}(\mathbf{\theta}_{n}+\mu\nabla_{\theta}f(\mathbf{\theta}_{n},\mathbf{Q}_{n}))$}
\State{$\mathbf{Q}_{n+1}=P_{\mathbf{Q}}(\mathbf{Q}_{n}+\mu\nabla_{\mathbf{Q}}f(\mathbf{\theta}_{n},\mathbf{Q}_{n}))$}
\EndFor
\end{algorithmic}
\end{algorithm}
\textcolor{black}{For any first-order method, exploiting the structure
of the optimization problem is key to speed up its convergence. In
this regard, we remark that the effective channel via the RIS is $\sqrt{\beta_{\mathrm{INDIR}}^{-1}}\mathbf{H}_{2}\mathbf{F}(\boldsymbol{\theta})\mathbf{\bar{\mathbf{H}}}_{1}$
which can be some orders of magnitude weaker or stronger than the
direct link $\bar{\mathbf{H}}_{\mathrm{DIR}}$. This unbalanced data
in \eqref{eq:capacityprob} makes Algorithm \ref{alg:GPA} converge
slowly.}

\textcolor{black}{To increase the convergence speed of Algorithm \ref{alg:GPA},
we propose a change of variable as follows:\begin{subequations}
\begin{align}
\bar{\mathbf{Q}} & =k^{2}\mathbf{Q}\label{eq:Q_scal}\\
\bar{\boldsymbol{\theta}} & =\boldsymbol{\theta}/k\label{eq:theta_scal}\\
\bar{\mathbf{H}}_{\mathrm{DIR}} & =\mathbf{H}_{\mathrm{DIR}}/(k\sqrt{N_{0}})\label{eq:Hdir_scale}
\end{align}
\end{subequations}for some $k>0$.  Accordingly, the equivalent
optimization problem with respect to the new variables $\bar{\boldsymbol{\theta}}$
and $\bar{\mathbf{Q}}$ reads\begin{subequations}\label{eq:capacityprob:scale}
\begin{align}
\underset{\bar{\boldsymbol{\theta}},\bar{\mathbf{Q}}}{\maximize} & \ f(\bar{\boldsymbol{\theta}},\bar{\mathbf{Q}})=\ln\det\left(\mathbf{I}+\mathbf{Z}(\bar{\boldsymbol{\theta}})\bar{\mathbf{Q}}\mathbf{Z}^{H}(\bar{\boldsymbol{\theta}})\right)\label{eq:opt_problem-1}\\
\st & \ \tr(\bar{\mathbf{Q}})\le\bar{P}_{t};\bar{\mathbf{Q}}\succeq\mathbf{0}\label{eq:pow_con_scale}\\
 & \ \bigl|\bar{\theta}_{l}\bigr|=\frac{1}{k},l=1,2,\ldots,N_{\mathrm{ris}}\label{eq:RIS_elem_con_scale}
\end{align}
\end{subequations}where $\bar{P}_{t}=k^{2}P_{t}$. The above change
of variable step is equivalent to scaling the gradient of the original
objective, which can improve the convergence rate. Now we solve \eqref{eq:capacityprob:scale}
following the iterative procedure in Algorithm \ref{alg:GPA}, but
instead of $\mathbf{Q}$ and $\boldsymbol{\theta}$ we use their scaled
versions $\bar{\mathbf{Q}}$ and $\bar{\boldsymbol{\theta}}$ which
are defined by the expressions \eqref{eq:Q_scal} and \eqref{eq:theta_scal},
respectively. Accordingly, the sets that contain all valid $\bar{\mathbf{Q}}$
and $\bar{\boldsymbol{\theta}}$ are denoted as $\bar{\Theta}$ and
$\bar{Q}$, respectively. The projections of computed $\bar{\mathbf{Q}}$
and $\bar{\boldsymbol{\theta}}$ onto $\bar{\Theta}$ and $\bar{Q}$
are performed in the same way as the projections in the previous subsections,
and the constraints \eqref{eq:pow_con} and \eqref{eq:RIS_elem_cons}
are replaced by \eqref{eq:pow_con_scale} and \eqref{eq:RIS_elem_con_scale},
respectively. Also, it should be pointed out that $\bar{\mathbf{H}}_{\mathrm{DIR}}$
is scaled (i.e., divided by $k$) in \eqref{eq:Hdir_scale} compared
to \eqref{eq:Hdir}. An appropriate value for $k$ should reflect
the difference between the direct and indirect links. When the direct
link is absent, $k$ should take into account the difference between
the feasible sets of $\mathbf{Q}$ and $\boldsymbol{\theta}$. From
our extensive numerical experiments, an appropriate value for $k$,
depending on the presence or absence of the direct link, is given
by}\footnote{\textcolor{black}{Since the gradients with respect to $\mathbf{Q}$
and $\boldsymbol{\theta}$ are of different sizes, using two different
step sizes for each of them can actually increase the convergence
rate of PGM. In order to preserve the single step size, we introduce
the scaling factor $k$ which provides the same effect as if two independent
step sizes were used. The value of the scaling factor $k$ is obtained
in a heuristic manner, by performing numerical experiments.}}\textcolor{black}{{} }%
\begin{comment}
\textcolor{black}{it is more natural to equivalently rewrite the equation
below in terms of scaled channels $\bar{\mathbf{H}}_{\mathrm{DIR}}$
and $\bar{\mathbf{H}}_{1}$}
\end{comment}
\textcolor{black}{
\begin{equation}
k=\begin{cases}
10\max\{1,\frac{1}{\sqrt{P_{t}}}\}\sqrt{\frac{1}{\beta_{\mathrm{INDIR}}^{-1/2}}\frac{\left\Vert \mathbf{H}_{\mathrm{DIR}}\right\Vert }{\left\Vert \mathbf{H}_{2}\mathbf{H}_{1}\right\Vert }}, & \mathbf{H}_{\mathrm{DIR}}\neq\mathbf{0}\\
10, & \mathbf{H}_{\mathrm{DIR}}=\mathbf{0}.
\end{cases}
\end{equation}
Based on the previous expressions, we can see that the PGM requires
only the knowledge of the }\textcolor{black}{\emph{cascaded}}\textcolor{black}{{}
channel, and not the individual channels $\mathbf{H}_{1}$ and $\mathbf{H}_{2}$,
for the indirect link. Estimation of this channel can be performed}\textit{\textcolor{black}{\emph{
at the receiver or the transmitter in TDD mode}}}\textcolor{black}{{}
\cite{lin2020reconfigurable,zappone2020overhead}.}

\section{\textcolor{black}{Convergence and Complexity Analysis\label{sec:Convergence-and-Complexity}}}

\subsection{\textcolor{black}{Convergence Analysis}}

\begin{figure*}[t]
\begin{comment}
I just use \textbackslash Vert to rewrite the equation to make it
look nicer and save some space.
\end{comment}
\begin{equation}
\left(\bigl\Vert\nabla_{\mathbf{\bar{\boldsymbol{\theta}}}}f(\bar{\boldsymbol{\theta}}_{1},\bar{\mathbf{Q}}_{1})-\nabla_{\mathbf{\bar{\boldsymbol{\theta}}}}f(\bar{\boldsymbol{\theta}}_{2},\bar{\mathbf{Q}}_{2})\bigr\Vert^{2}+\bigl\Vert\nabla_{\bar{\mathbf{Q}}}f(\bar{\boldsymbol{\theta}}_{1},\bar{\mathbf{Q}}_{1})-\nabla_{\bar{\mathbf{Q}}}f(\bar{\boldsymbol{\theta}}_{2},\bar{\mathbf{Q}}_{2})\bigr\Vert^{2}\right)^{1/2}\leq L\left(\bigl\Vert\bar{\mathbf{Q}}_{1}-\bar{\mathbf{Q}}_{2}\bigr\Vert^{2}+\bigl\Vert\bar{\boldsymbol{\theta}}_{1}-\bar{\boldsymbol{\theta}}_{2}\bigr\Vert^{2}\right)^{1/2}\label{eq:LIP_def}
\end{equation}
\end{figure*}
\textcolor{black}{In this subsection we prove the convergence of Algorithm
\ref{alg:GPA} for solving \eqref{eq:capacityprob:scale}, following
the framework in \cite{Li:2015:APG}. To achieve this, we first show
that $f(\bar{\boldsymbol{\theta}},\bar{\mathbf{Q}})$ has a Lipschitz
continuous gradient with a Lipschitz constant $L$, and then assert
that Algorithm \ref{alg:GPA} is convergent if the step size satisfies
$\mu\leq\frac{1}{L}$. For the first part of the proof, recall that
a function $f(\mathbf{x})$ is said to be $L$-Lipschitz continuous
(also known as $L$-smooth) over a set $\mathcal{X}$ if for all $\mathbf{x},\mathbf{y}\in\mathcal{X}$
we have
\begin{equation}
||\nabla f(\mathbf{x})-\nabla f(\mathbf{y})||\leq L||\mathbf{x}-\mathbf{y}||.\label{eq:LIP_orig}
\end{equation}
 }%
\begin{comment}
\textcolor{black}{perhaps we can use \textbackslash Vert to write
the norms throughout the paper, instead of || ||}
\end{comment}
\textcolor{black}{In our present context, the inequality \eqref{eq:LIP_orig}
corresponds to \eqref{eq:LIP_def}, to prove which we may make use
of the following lemma.}
\begin{lem}
\textcolor{black}{\label{lem:Lipschitz:part}The following inequalities
hold for $\nabla_{\mathbf{\bar{\boldsymbol{\theta}}}}f(\bar{\boldsymbol{\theta}},\bar{\mathbf{Q}})$
and $\nabla_{\bar{\mathbf{Q}}}f(\bar{\boldsymbol{\theta}},\bar{\mathbf{Q}})$
\begin{multline}
||\nabla_{\mathbf{\bar{\boldsymbol{\theta}}}}f(\bar{\boldsymbol{\theta}}_{1},\bar{\mathbf{Q}}_{1})-\nabla_{\mathbf{\bar{\boldsymbol{\theta}}}}f(\bar{\boldsymbol{\theta}}_{2},\bar{\mathbf{Q}}_{2})||\\
\le(ab+ab^{3}\bar{P}_{t})||\bar{\mathbf{Q}}_{1}-\bar{\mathbf{Q}}_{2}||+(a^{2}\bar{P}_{t}+2a^{2}b^{2}\bar{P}_{t}^{2})||\bar{\boldsymbol{\theta}}_{1}-\bar{\boldsymbol{\theta}}_{2}||\label{eq:grad_phi_dif_fin}
\end{multline}
\begin{multline}
||\nabla_{\bar{\mathbf{Q}}}f(\bar{\boldsymbol{\theta}}_{1},\bar{\mathbf{Q}}_{1})-\nabla_{\bar{\mathbf{Q}}}f(\bar{\boldsymbol{\theta}}_{2},\bar{\mathbf{Q}}_{2})||\\
\le b^{4}||\bar{\mathbf{Q}}_{1}-\bar{\mathbf{Q}}_{2}||+(2ab+2ab^{3}\bar{P}_{t})||\bar{\boldsymbol{\theta}}_{1}-\bar{\boldsymbol{\theta}}_{2}||\label{eq:grad_Q_dif_fin}
\end{multline}
}where 
\begin{align}
a & =\lambda_{\max}(\bar{\mathbf{H}}_{1})\lambda_{\max}(\mathbf{H}_{2})\\
b & =\lambda_{\max}(\mathbf{\bar{\mathbf{H}}}_{\mathrm{DIR}})+k^{-1}\lambda_{\max}(\bar{\mathbf{H}}_{1})\lambda_{\max}(\mathbf{H}_{2}).
\end{align}
\end{lem}
\begin{IEEEproof}
See Appendix \ref{sec:Proof:lipschitz:part}.
\end{IEEEproof}
\textcolor{black}{With the aid of the above lemma, we assert the smoothness
of $f(\bar{\boldsymbol{\theta}},\bar{\mathbf{Q}})$ in the next theorem.}%
\begin{comment}
\textcolor{black}{Now we should be consistent with Lemma or Theorem
and use a proper environment.}
\end{comment}
\setcounter{thm}{0}
\begin{thm}
\textcolor{black}{\label{thm:Lipschitz:obj} The objective $f(\bar{\boldsymbol{\theta}},\bar{\mathbf{Q}})$
is $L$-smooth with a constant $L$ given by
\begin{gather}
L=\sqrt{\max(L_{\bar{\boldsymbol{\theta}}}^{2},L_{\bar{\mathbf{Q}}}^{2})},\label{eq:Lip_val}
\end{gather}
where}
\begin{align}
L_{\bar{\boldsymbol{\theta}}}^{2}= & (2ab^{5}+4a^{2}b^{2})+(a^{3}b+2ab^{7}+8a^{2}b^{4})\bar{P}_{t}\nonumber \\
 & +(3a^{3}b^{3}+a^{4}+4a^{2}b^{6})\bar{P}_{t}^{2}+(2a^{3}b^{5}+4a^{4}b^{2})\bar{P}_{t}^{3}\nonumber \\
 & +4a^{4}b^{4}\bar{P}_{t}^{4}\\
L_{\bar{\mathbf{Q}}}^{2}= & (a^{2}b^{2}+b^{8}+2ab^{5})+(2a^{2}b^{4}+a^{3}b+2ab^{7})\bar{P}_{t}\nonumber \\
 & +(a^{2}b^{6}+3a^{3}b^{3})\bar{P}_{t}^{2}+2a^{3}b^{5}\bar{P}_{t}^{3}
\end{align}
\end{thm}
\begin{IEEEproof}
Theorem \ref{thm:Lipschitz:obj} follows immediately from Lemma \ref{lem:Lipschitz:part}
and the inequality
\[
2||\bar{\mathbf{Q}}_{1}-\bar{\mathbf{Q}}_{2}||\times||\bar{\boldsymbol{\theta}}_{1}-\bar{\boldsymbol{\theta}}_{2}||\le||\bar{\mathbf{Q}}_{1}-\bar{\mathbf{Q}}_{2}||^{2}+||\bar{\boldsymbol{\theta}}_{1}-\bar{\boldsymbol{\theta}}_{2}||^{2}.
\]
Specifically, we have\textcolor{black}{
\begin{multline*}
||\nabla_{\mathbf{\bar{\boldsymbol{\theta}}}}f(\bar{\boldsymbol{\theta}}_{1},\bar{\mathbf{Q}}_{1})-\nabla_{\mathbf{\bar{\boldsymbol{\theta}}}}f(\bar{\boldsymbol{\theta}}_{2},\bar{\mathbf{Q}}_{2})||^{2}\\
+||\nabla_{\bar{\mathbf{Q}}}f(\bar{\boldsymbol{\theta}}_{1},\bar{\mathbf{Q}}_{1})-\nabla_{\bar{\mathbf{Q}}}f(\bar{\boldsymbol{\theta}}_{2},\bar{\mathbf{Q}}_{2})||^{2}\\
\le L_{\bar{\mathbf{Q}}}^{2}||\bar{\mathbf{Q}}_{1}-\bar{\mathbf{Q}}_{2}||^{2}+L_{\bar{\boldsymbol{\theta}}}^{2}||\bar{\boldsymbol{\theta}}_{1}-\bar{\boldsymbol{\theta}}_{2}||^{2}\\
\le\max(L_{\bar{\boldsymbol{\theta}}}^{2},L_{\bar{\mathbf{Q}}}^{2})\bigl(||\bar{\mathbf{Q}}_{1}-\bar{\mathbf{Q}}_{2}||^{2}+||\bar{\boldsymbol{\theta}}_{1}-\bar{\boldsymbol{\theta}}_{2}||^{2}\bigr).
\end{multline*}
Taking the square root of both sides of the above inequality, we can
see that $\sqrt{\max(L_{\bar{\boldsymbol{\theta}}}^{2},L_{\bar{\mathbf{Q}}}^{2})}$
is a Lipschitz constant of the gradient of $f(\bar{\boldsymbol{\theta}},\bar{\mathbf{Q}})$;
this completes the proof.}
\end{IEEEproof}
\textcolor{black}{The convergence of Algorithm \ref{alg:GPA} is stated
in the following theorem.}
\begin{thm}
\textcolor{black}{\label{thm:convergence}Assume the step size satisfies
$\mu<\frac{1}{L}$, where $L$ is given in \eqref{eq:Lip_val}. Then
the iterates $(\bar{\boldsymbol{\theta}}_{n},\bar{\mathbf{Q}}_{n})$
generated by Algorithm \ref{alg:GPA} are bounded. Let $(\boldsymbol{\theta}^{\ast},\mathbf{Q}^{\ast})$
be any accumulation point of the set $\{(\bar{\boldsymbol{\theta}}_{n},\bar{\mathbf{Q}}_{n})\}$,
then, $(\boldsymbol{\theta}^{\ast},\mathbf{Q}^{\ast})$ is a critical
point of \eqref{eq:capacityprob}.}
\end{thm}
\begin{IEEEproof}
\textcolor{black}{See Appendix \ref{sec:convergence}.}
\end{IEEEproof}
\textcolor{black}{Before proceeding further, a subtle point regarding
the convergence of Algorithm \ref{alg:GPA} is worth mentioning. Specifically,
the proposed method is provably convergent to a critical point of
the considered problem, also known as a stationary solution, which
satisfies the }\textcolor{black}{\emph{necessary}}\textcolor{black}{{}
optimality conditions for \eqref{eq:capacityprob}. However, since
\eqref{eq:capacityprob} is nonconvex, these optimality conditions
may not be sufficient in general, and thus the solution obtained from
Algorithm \ref{alg:GPA} may not be globally optimal. However, it
is often the case that with a good initialization, a stationary solution
is good enough for practical applications. We note that the same comments
also apply to the AO method proposed in \cite{zhang2019capacity}.
}

\subsection{\textcolor{black}{Complexity Analysis\label{subsec:PGM_comp}}}

\textcolor{black}{}%
\textcolor{black}{In this subsection, we analyze the computational
complexity of Algorithm \ref{alg:GPA} (i.e., the \ac{PGM})}.\footnote{Although the PGM is actually implemented in the scaled-variable form
described in Subsection \ref{subsec:Imp-Conv}, this scaling does
not affect the complexity of the PGM which is equal to the complexity
of Algorithm \ref{alg:GPA}. Therefore, we analyze the complexity
of Algorithm \ref{alg:GPA} in the sequel.}\textcolor{black}{{} To simplify the analysis while }still\textcolor{black}{{}
providing a good approximation to the complexity of Algorithm \ref{alg:GPA},
we concentrate on the number of complex multiplications required per
iteration. To this end, we first recall some fundamental results.
Specifically, the multiplication of $\mathbf{A}\in\mathbb{C}^{m\times n}$
and $\mathbf{B}\in\mathbb{C}^{n\times p}$ needs $mnp$ complex multiplications
when $\mathbf{A}$ and $\mathbf{B}$ are dense matrices.}\footnote{Special algorithms can reduce the complexity further, but this is
not our focus in this paper.}\textcolor{black}{{} This complexity reduces to $mn$ for the case of
a square diagonal matrix $\mathbf{B}\in\mathbb{C}^{n\times n}$. Calculating
$\vect_{d}(\mathbf{A}\mathbf{B}\mathbf{C})$, $\mathbf{A}\in\mathbb{C}^{m\times n}$,
$\mathbf{B}\in\mathbb{C}^{n\times p}$ and $\mathbf{C}\in\mathbb{C}^{p\times m}$,
needs $mp(n+1)$ complex multiplications, which is justified as follows.
Multiplying a row of $\mathbf{A}$ with $\mathbf{B}$ requires $np$
complex multiplications and multiplying the resulting row vector with
the corresponding column of $\mathbf{C}$ requires a further $p$
}complex\textcolor{black}{{} multiplications.}

\textcolor{black}{It is obvious that the complexity of the proposed
method is determined by Steps 3 and 4 in Algorithm \ref{alg:GPA}.
The computation of $\mathbf{Z}(\boldsymbol{\theta})$ is dominated
by that of the term $\mathbf{H}_{2}\mathbf{F}(\boldsymbol{\theta})\bar{\mathbf{H}}_{1}$,
which requires $N_{r}N_{\mathrm{ris}}+N_{r}N_{t}N_{\mathrm{ris}}$
complex multiplications. To compute $\nabla_{\boldsymbol{\theta}}f(\boldsymbol{\theta},\mathbf{Q})$,
we also need to compute the term $\mathbf{A}=\mathbf{K}(\boldsymbol{\theta},\mathbf{Q})\mathbf{Z}(\boldsymbol{\theta})\in\mathbb{C}^{N_{r}\times N_{t}}$.
Instead of directly computing $\mathbf{K}(\boldsymbol{\theta},\mathbf{Q})=(\mathbf{I}+\mathbf{Z}(\boldsymbol{\theta})\mathbf{Q}\mathbf{Z}^{H}(\boldsymbol{\theta}))^{-1}$
using matrix inversion and then multiplying $\mathbf{K}(\boldsymbol{\theta},\mathbf{Q})$
with $\mathbf{Z}(\boldsymbol{\theta})$, we note that $\mathbf{A}$
is in fact the solution to the linear system $\bigl(\mathbf{I}+\mathbf{Z}(\boldsymbol{\theta})\mathbf{Q}\mathbf{Z}^{H}(\boldsymbol{\theta})\bigr)\mathbf{X}=\mathbf{Z}(\boldsymbol{\theta})$.
To form $\mathbf{Z}(\boldsymbol{\theta})\mathbf{Q}\mathbf{Z}^{H}(\boldsymbol{\theta})$,
we first need $N_{r}N_{t}^{2}$ multiplications to achieve $\mathbf{Z}(\boldsymbol{\theta})\mathbf{Q}$
and then $(N_{r}^{2}+N_{r})N_{t}/2$ to multiply $\mathbf{Z}(\boldsymbol{\theta})\mathbf{Q}$
with $\mathbf{Z}^{H}(\boldsymbol{\theta})$. Solving the linear system
using Cholesky decomposition, by solving two triangular systems using
forward and backward substitution, requires a complexity which is
$\mathcal{O}(N_{r}^{3}+N_{r}^{2}N_{t})$. In summary, the computation
of $\mathbf{A}$ takes $\mathcal{O}(N_{t}^{2}N_{r}+\frac{3}{2}N_{t}N_{r}^{2}+N_{r}^{3})$
multiplications. Next, the computation of $\mathbf{A}\mathbf{Q}$
requires $N_{r}N_{t}^{2}$ complex multiplications. To calculate $\vect_{d}(\mathbf{H}_{2}^{H}\mathbf{A}\mathbf{Q}\bar{\mathbf{H}}_{1}^{H})$,
we need $N_{\mathrm{ris}}N_{t}(N_{r}+1)$ complex multiplications.
As $\mathbf{A}$ is also common to \eqref{eq:grad_Q}, the complexity
of computing $\nabla_{\mathbf{Q}}f(\boldsymbol{\theta},\mathbf{Q})$
is only $N_{r}N_{t}^{2}$. In summary, the computational complexity
of $\nabla_{\boldsymbol{\theta}}f(\boldsymbol{\theta},\mathbf{Q})$
and $\nabla_{\mathbf{Q}}f(\boldsymbol{\theta},\mathbf{Q})$ is $\mathcal{O}\bigl(2N_{\mathrm{ris}}N_{t}N_{r}+2N_{t}^{2}N_{r}+\frac{3}{2}N_{t}N_{r}^{2}+N_{r}^{3}+N_{r}N_{\mathrm{ris}}+N_{t}N_{\mathrm{ris}}\bigr)$.
When $N_{\mathrm{ris}}$ is much larger than $N_{t}$ and $N_{r}$,
then the complexity can be approximated by $\mathcal{O}\bigl(N_{\mathrm{ris}}N_{t}N_{r})$.}

\textcolor{black}{}
\begin{table}[t]
\textcolor{black}{\caption{Comparison of the computational complexity required by the proposed
PGM method and the AO method to reach 95\,\% of the average achievable
rate at the 500th iteration.\label{tab:Comp}}
}
\centering{}\textcolor{black}{}%
\begin{tabular}{ccccccc}
\toprule 
\textcolor{black}{\footnotesize{}Direct link} & \textcolor{black}{\footnotesize{}$N_{\mathrm{ris}}$} & \textcolor{black}{\footnotesize{}$I_{\mathrm{PGM}}$} & \textcolor{black}{\footnotesize{}$C_{\mathrm{PGM,IT}}$} & \textcolor{black}{\footnotesize{}$C_{\mathrm{PGM}}$} & \textcolor{black}{\footnotesize{}$I_{\mathrm{OI}}$} & \textcolor{black}{\footnotesize{}$C_{\mathrm{AO}}$}\tabularnewline
\midrule
\multirow{4}{*}{\textcolor{black}{\footnotesize{}Present}} & \textcolor{black}{\footnotesize{}100} & \textcolor{black}{\footnotesize{}19} & \textcolor{black}{\footnotesize{}9436} & \textcolor{black}{\footnotesize{}179284} & \textcolor{black}{\footnotesize{}1} & \textcolor{black}{\footnotesize{}394304}\tabularnewline
 & \textcolor{black}{\footnotesize{}225} & \textcolor{black}{\footnotesize{}6} & \textcolor{black}{\footnotesize{}19311} & \textcolor{black}{\footnotesize{}115866} & \textcolor{black}{\footnotesize{}1} & \textcolor{black}{\footnotesize{}862304}\tabularnewline
 & \textcolor{black}{\footnotesize{}400} & \textcolor{black}{\footnotesize{}4} & \textcolor{black}{\footnotesize{}33136} & \textcolor{black}{\footnotesize{}132544} & \textcolor{black}{\footnotesize{}1} & \textcolor{black}{\footnotesize{}1517504}\tabularnewline
 & \textcolor{black}{\footnotesize{}625} & \textcolor{black}{\footnotesize{}3} & \textcolor{black}{\footnotesize{}50911} & \textcolor{black}{\footnotesize{}152733} & \textcolor{black}{\footnotesize{}1} & \textcolor{black}{\footnotesize{}2359904}\tabularnewline
\midrule 
\multirow{4}{*}{\textcolor{black}{\footnotesize{}Blocked}} & \textcolor{black}{\footnotesize{}100} & \textcolor{black}{\footnotesize{}2} & \textcolor{black}{\footnotesize{}9436} & \textcolor{black}{\footnotesize{}18872} & \textcolor{black}{\footnotesize{}1} & \textcolor{black}{\footnotesize{}394304}\tabularnewline
 & \textcolor{black}{\footnotesize{}225} & \textcolor{black}{\footnotesize{}2} & \textcolor{black}{\footnotesize{}19311} & \textcolor{black}{\footnotesize{}38622} & \textcolor{black}{\footnotesize{}1} & \textcolor{black}{\footnotesize{}862304}\tabularnewline
 & \textcolor{black}{\footnotesize{}400} & \textcolor{black}{\footnotesize{}2} & \textcolor{black}{\footnotesize{}33136} & \textcolor{black}{\footnotesize{}66272} & \textcolor{black}{\footnotesize{}1} & \textcolor{black}{\footnotesize{}1517504}\tabularnewline
 & \textcolor{black}{\footnotesize{}625} & \textcolor{black}{\footnotesize{}2} & \textcolor{black}{\footnotesize{}50911} & \textcolor{black}{\footnotesize{}101822} & \textcolor{black}{\footnotesize{}1} & \textcolor{black}{\footnotesize{}2359904}\tabularnewline
\bottomrule
\end{tabular}
\end{table}
\textcolor{black}{Next, multiplying $\mu$ with $\nabla_{\boldsymbol{\theta}}f(\boldsymbol{\theta},\mathbf{Q})$
and then projecting the result onto $\Theta$ requires $3N_{\mathrm{ris}}$
complex multiplications. Similarly, we need $N_{t}^{2}/2$ operations
to multiply $\mu$ with $\nabla_{\mathbf{Q}}f(\boldsymbol{\theta}_{n},\mathbf{Q}_{n})$.
The projection of $\mathbf{Q}_{n}+\mu\nabla_{\mathbf{Q}}f(\boldsymbol{\theta}_{n},\mathbf{Q}_{n})$
onto $\mathcal{Q}$ requires: $\mathcal{O}(N_{t}^{3})$ operations
for the eigenvalue decomposition, $\mathcal{O}(N_{t}^{2})$ operations
for the water-filling algorithm in \eqref{eq:WF} and $N_{t}^{2}+(N_{t}^{2}+N_{t})N_{t}/2$
operations for the matrix multiplication $\mathbf{Q}=\mathbf{UD}\mathbf{U}^{H}$.
Therefore, the complexity for the update and projection operations
in Step 4 is given by $\mathcal{O}(\frac{3}{2}N_{t}^{3})$.}%
\begin{comment}
\textcolor{black}{If we ignore the terms with lower order, we need
to be consistent throughout this section. That means, the analysis
looks much simpler. For example we can ignore the term $N_{r}N_{\mathrm{ris}}+N_{t}N_{\mathrm{ris}}$
in the complexity of the gradient above since $N_{\mathrm{ris}}N_{t}N_{r}$
is already there.}
\end{comment}
\textcolor{black}{{} Thus, the per-iteration complexity of Algorithm
\ref{alg:GPA} is finally determined as
\begin{align}
C_{\mathrm{PGM,IT}}=\mathcal{O}(2N_{\mathrm{ris}}N_{t}N_{r}+2N_{t}^{2}N_{r}+\frac{3}{2}N_{t}N_{r}^{2}+N_{r}^{3}\nonumber \\
+N_{r}N_{\mathrm{ris}}+N_{t}N_{\mathrm{ris}}+3N_{\mathrm{ris}}+\frac{3}{2}N_{t}^{3}),\label{eq:pgm_comp}
\end{align}
while the total complexity $C_{\mathrm{PGM}}$ also depends from the
number of required iterations $I_{\mathrm{PGM}}$. The computational
complexity of the proposed PGM and the AO method from \cite{zhang2019capacity}
are presented in Table \ref{tab:Comp}. The complexity of the AO is
expressed with respect to the number of outer iterations $I_{\mathrm{OI}}$,
where one outer iteration is actually a sequence of $N_{\mathrm{ris}}+1$
conventional iterations. Further details and discussion on this complexity
comparison will be presented in Subsection \ref{subsec:Comp-compl-results}.}

\subsection{\textcolor{black}{Improved Convergence by Backtracking Line Search}}

\textcolor{black}{It often occurs that the Lipschitz constant given
in \eqref{eq:Lip_val} is much larger than the best Lipschitz constant
for the gradient of the objective. The corresponding step size required
(according to Theorem \ref{thm:convergence}) to guarantee the convergence
is then very small, and adopting this step size can lead to a very
slow convergence. To speed up the convergence of the proposed PGM,
we can employ a backtracking line search to find a possibly larger
step size at each iteration. In the following, we present a line search
procedure, based on the Armijo\textendash Goldstein condition \cite{armijo1966minimization},
that is numerically shown to be efficient for our considered problem.}

\textcolor{black}{Let $L_{0}>0$, $\delta>0$ be a small constant,
and $\rho\in(0,1)$. In Steps 3 and 4 of Algorithm \ref{alg:GPA},
we replace the step size $\mu$ by $L_{o}\rho^{k_{n}}$ and obtain
}\eqref{eq:Theta_search} and \eqref{eq:Q_search}\textcolor{black}{,
where $k_{n}$ is the smallest nonnegative integer that satisfies
}\eqref{eq:ls-cond}\textcolor{black}{.}\begin{subequations}\label{eq:linesearch}
\begin{align}
\boldsymbol{\theta}_{n+1} & =P_{\Theta}(\boldsymbol{\theta}_{n}+L_{o}\rho^{k_{n}}\nabla_{\boldsymbol{\theta}}f(\boldsymbol{\theta}_{n},\mathbf{Q}_{n}))\label{eq:Theta_search}\\
\mathbf{Q}_{n+1} & =P_{\mathcal{Q}}(\mathbf{Q}_{n}+L_{o}\rho^{k_{n}}\nabla_{\mathbf{Q}}f(\boldsymbol{\theta}_{n},\mathbf{Q}_{n}))\label{eq:Q_search}\\
f(\boldsymbol{\theta}_{n+1},\mathbf{Q}_{n+1}) & \geq f(\boldsymbol{\theta}_{n},\mathbf{Q}_{n})+\nonumber \\
 & \delta\bigl(||\boldsymbol{\theta}_{n+1}-\boldsymbol{\theta}_{n}||^{2}+||\mathbf{Q}_{n+1}-\mathbf{Q}_{n}||^{2}\bigr).\label{eq:ls-cond}
\end{align}
\end{subequations}

\textcolor{black}{The }above\textcolor{black}{{} backtracking line search
can be found through an iterative procedure, which is guaranteed to
}terminate\textcolor{black}{{} after a finite number of iterations since
$f(\bar{\boldsymbol{\theta}},\bar{\mathbf{Q}})$ is $L$-smooth. It
is easy to see that the convergence of Algorithm \ref{alg:GPA} (i.e.,
Theorem \ref{thm:convergence}) }still\textcolor{black}{{} holds when
this procedure is used to find the step size. We remark that the line
search described above results in increased per-iteration complexity.
Suppose that the line search stops after $I_{\mathrm{LS}}$ steps,
the additional complexity is $\mathcal{O}\bigl(I_{\mathrm{LS}}(3N_{\mathrm{ris}}+2N_{t}^{3})\bigr)$.
However, this computational cost }turns out to be\textcolor{black}{{}
 immaterial, since the line search can significantly reduce the required
number of iterations and hence the actual overall run time.}

\section{\textcolor{black}{Total FPSL Ratio - A Metric of RIS Applicability\label{sec:Total-FSPL}}}

\textcolor{black}{It }can be\textcolor{black}{{} very useful to have
a first-order estimate of the benefit }(if any)\textcolor{black}{{}
provided to a wireless communication system by adding an RIS. We can
achieve this by considering the total FSPL of the indirect and direct
links. Note that when speaking about the ``total FSPL'' of the indirect
link, we require (in contrast to \eqref{eq:FSPL_indir}) a definition
which takes into account also the RIS phase shift~values.}

\textcolor{black}{The computation of the total FSPL of the indirect
link is an intractable problem in a MIMO system, since the optimal
RIS element phase shifts are }\textcolor{black}{\emph{a priori}}\textcolor{black}{{}
unknown and can only be obtained by implementing an iterative optimization
method. This problem was approximately tackled only for the single-stream
scenario in \cite{zappone2020overhead}, and the obtained results
were then used in \cite{qian2020beamforming} to quantify the performance
of RISs in the far field regime} (which is also the case considered
in this paper)\textcolor{black}{, but only for Rayleigh and deterministic
\ac{LOS} channels. To overcome this issue, we consider the total
FSPL of the indirect link in a \ac{SISO} system, which is given by
\begin{equation}
\beta_{\mathrm{INDIR,T}}^{-1}=\beta_{\mathrm{INDIR}}^{-1}\mathbb{E}\left\{ \left|\mathbf{h}_{2}\mathbf{F}(\boldsymbol{\theta})\mathbf{h}_{1}\right|^{2}\right\} ,\label{eq:beta-indir-1}
\end{equation}
where $\mathbf{h}_{1}$ models the channel between the transmit antenna
and the RIS, and $\mathbf{h}_{2}$ models the channel between the
RIS and the receive antenna. The optimal RIS element phase shift values
in \eqref{eq:beta-indir-1} satisfy $\phi_{i}=-\arg\left\{ h_{2}(i)h_{1}(i)\right\} $
and as a result we have $\mathbf{h}_{2}\mathbf{F}(\boldsymbol{\theta})\mathbf{h}_{1}=\sum_{i=1}^{N_{\mathrm{ris}}}\left|h_{2}(i)h_{1}(i)\right|$.
As all of the terms $\left|h_{2}(i)h_{1}(i)\right|$ follow the same
distribution, we may write
\begin{equation}
\mathbb{E}\{\sum_{i=1}^{N_{\mathrm{ris}}}\left|h_{2}(i)h_{1}(i)\right|\}=N_{\mathrm{ris}}\mathbb{E}\left\{ \left|h_{2}(1)h_{1}(1)\right|\right\} .
\end{equation}
From Jensen's inequality we obtain
\begin{multline}
\mathbb{E}\left\{ \left|\mathbf{h}_{2}\mathbf{F}(\boldsymbol{\theta})\mathbf{h}_{1}\right|^{2}\right\} =\mathbb{E}\left\{ \left|\sum_{i=1}^{N_{\mathrm{ris}}}\left|h_{2}(i)h_{1}(i)\right|\right|^{2}\right\} \ge\\
\left(\mathbb{E}\left\{ \sum_{i=1}^{N_{\mathrm{ris}}}\left|h_{2}(i)h_{1}(i)\right|\right\} \right)^{2}=N_{\mathrm{ris}}^{2}\left(\mathbb{E}\left\{ \left|h_{2}(1)h_{1}(1)\right|\right\} \right)^{2}.\label{eq:pow_equ}
\end{multline}
Substituting \eqref{eq:pow_equ} into \eqref{eq:beta-indir-1}, we
finally obtain
\begin{equation}
\beta_{\mathrm{INDIR,T}}^{-1}\ge\beta_{\mathrm{INDIR}}^{-1}N_{\mathrm{ris}}^{2}\left(\mathbb{E}\left\{ \left|h_{2}(1)h_{1}(1)\right|\right\} \right)^{2},\label{eq:Total_FPSL_indir}
\end{equation}
which constitutes an upper-bound on the total FSPL. The total FSPL
of the direct link in a \ac{SISO} system is given by $\beta_{\mathrm{DIR,T}}=\beta_{\mathrm{DIR}}$,
since the direct link does not alter the average signal power. Finally,
the ratio between the total \ac{FSPL} of the indirect and direct
links can be expressed as
\begin{equation}
T=\frac{\beta_{\mathrm{INDIR,T}}}{\beta_{\mathrm{DIR,T}}}=\frac{16}{\lambda^{2}}\frac{(d_{1}d_{2})^{2}}{d_{0}^{\alpha_{\mathrm{DIR}}}}\frac{1}{(l_{t}/d_{1}+l_{r}/d_{2})^{2}N_{\mathrm{ris}}^{2}E},\label{eq:FSPL_ratio-1-1}
\end{equation}
where $E=\left(\mathbb{E}\left\{ \left|h_{2}(1)h_{1}(1)\right|\right\} \right)^{2}$.
The obtained $T$ serves as a first-order measure of the applicability
of an RIS for a given communication scenario}\footnote{\textcolor{black}{Since \eqref{eq:FSPL_ratio-1-1} is derived for
a SISO system, $T$ is realistically only a rough measure of the applicability
of an RIS in a MIMO system. However, as we shall see in Section V,
this metric is quite useful for MIMO scenarios.}}\textcolor{black}{. For $T>1$, the direct link is }expected to be
always\textcolor{black}{{} stronger than the indirect link, even if
the RIS phase shifts are optimally adjusted. Consequently, the RIS
is capable of achieving limited performance gains with respect to
the case when only the direct link is utilized for communication.
For $T<1$, the indirect link with the optimized RIS phase shifts
is stronger than the direct link and the gains of using the RIS are
}usually more\textcolor{black}{{} substantial.}

\section{\textcolor{black}{Simulation Results\label{sec:Simulation-Results}}}

\textcolor{black}{In this section, we evaluate the achievable rate
of the proposed optimization algorithm with the aid of Monte Carlo
simulations. First, the study is conducted for a typical outdoor propagation
environment in two different scenarios: with the direct link present
and with the direct link blocked. For the case study where the direct
link is present, we utilize three benchmark schemes. The first benchmark
scheme is based on the implementation of the AO method from \cite{zhang2019capacity}.
The second and third benchmark schemes are based on the use of the
PGM in the case where only the indirect link is active and where only
the direct link is active, respectively. In the case where the direct
link is blocked, we only consider the AO method as the benchmark scheme.
Additionally, we show the variation of the achievable rate with the
number of RIS elements. Furthermore, we study the suitability of RIS-aided
wireless communications (with the proposed optimization method) for
implementation in indoor propagation environments. We also present
a comparison of the proposed and benchmark schemes in terms of computational
complexity and run time. In addition, we analyze the sensitivity and
robustness of the proposed PGM. Finally, we evaluate the influence
of data scaling and the line search procedure.}

\textcolor{black}{In the following simulation setup, the parameters
are $f=2\,\mathrm{GHz}$ (i.e., $\lambda=15\,\mathrm{cm}$), $s_{t}=s_{r}=\lambda/2=7.5\,\mathrm{cm}$,
$s_{\mathrm{ris}}=\lambda/2=7.5\,\mathrm{cm}$, $D=500\,\mathrm{m}$,
$N_{t}=8$, $N_{r}=4$, $\alpha_{\mathrm{DIR}}=3$, $N_{\mathrm{ris}}=225$,
$K=1$, $P_{t}=0\,\mathrm{dB}$ and $N_{0}=-120\thinspace\mathrm{dB}$.
The RIS elements are placed in a $15\times15$ square formation so
that the area of the RIS is slightly larger than $1\thinspace\mathrm{m}^{2}$.
The line search procedure for the proposed gradient algorithms utilizes
the parameters $L_{0}=10^{4}$, $\delta=10^{-5}$ and $\rho=1/2$.
Also, the minimum allowed step size value is the largest step size
value lower than $10^{-4}$. Unless otherwise specified, we assume
the initial values $\boldsymbol{\theta}=[1\;1\;\cdots\;1]^{T}$ and
$\mathbf{Q}=(P_{t}/N_{t})\mathbf{I}$ for all optimization algorithms.
To maintain compatibility with \cite{zhang2019capacity}, we set the
number of random initializations for the AO to $L_{\mathrm{AO}}=100$.
All of the achievable rate results, except those for very large $N_{\mathrm{ris}}$
in Figs. \ref{fig:RateVsNris} and \ref{fig:Achievable-rate_vs_freq},
are averaged over 200 independent channel realizations.}

\subsection{\textcolor{black}{Achievable Rate in Outdoor Environments}}

\subsubsection{\textcolor{black}{Direct link present}}

In this subsection, w\textcolor{black}{e present the achievable rate
simulation results when the considered communication system is located
in an outdoor environment. To obtain a more complete picture, the
positions of the transmitter and the receiver, as well as the position
of the RIS, are varied in simulations. In general, we analyze two
cases for the transmitter and receiver positions: the transmitter
and receiver are at substantially different distances from the plane
containing the RIS ($l_{t}\neq l_{r}$), and the transmitter and receiver
are at the same distance from the plane containing the RIS ($l_{t}=l_{r}$).
In each of these cases, the position of the RIS is also varied.}

\textcolor{black}{}
\begin{figure}[t]
\begin{centering}
\textcolor{black}{}\subfloat[The total FSPL ratio ($T$) versus $d_{\mathrm{ris}}$.\label{fig:Diff_distances_FSPL}]{\centering{}\textcolor{black}{\includegraphics[scale=0.8]{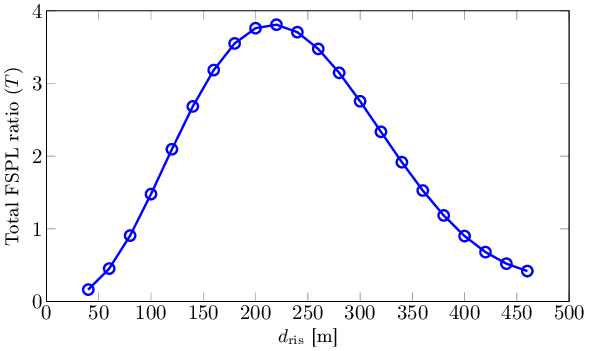}}}\textcolor{black}{\hfill{}}\subfloat[The total indirect link length $(d_{1}+d_{2})$ versus $d_{\mathrm{ris}}$.\label{fig:Total-ind-link-len}]{\centering{}\textcolor{black}{\includegraphics[scale=0.8]{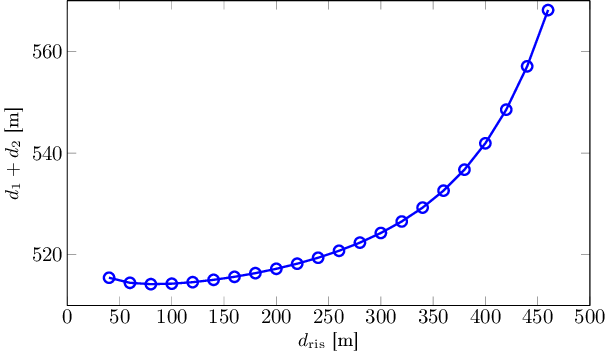}}}
\par\end{centering}
\centering{}\textcolor{black}{\caption{The total FSPL ratio and the indirect link length for $l_{t}=20\,\mathrm{m}$
and $l_{r}=100\,\mathrm{m}$. \label{fig:Diff_distances}}
}
\end{figure}
\textcolor{black}{In the first case, we assume $l_{t}=20\,\mathrm{m}$
and $l_{r}=100\,\mathrm{m}$. The variation of the FSPL ratio $T$
given by \eqref{eq:FSPL_ratio-1-1} and the total indirect link length
with the RIS position are shown in Fig. \ref{fig:Diff_distances}.
We observe that the highest FSPL ratio $T$ is obtained when the RIS
is placed close to the center, and the lowest $T$ is obtained when
the RIS is placed close to the transmitter or the receiver. In other
words, placing the RIS in the vicinity of the transmitter or the receiver
ensures the lowest signal attenuation for the indirect communication
link. It is interesting to note from Fig.~\ref{fig:Total-ind-link-len}
that in contrast to signal propagation principles for conventional
communication systems, the total length of the indirect link $(d_{1}+d_{2})$
does not determine the total FSPL of that link, and that the relationship
between these variables is not monotonic. For example, the minimum
and the maximum of the FSPL ratio $T$ in Fig. \ref{fig:Diff_distances_FSPL}
are obtained for almost the same total length of the indirect link,
as shown in Fig. \ref{fig:Total-ind-link-len}. Also, the largest
indirect link length in Fig. \ref{fig:Total-ind-link-len} does not
coincide with the highest FSPL of the indirect link in Fig. \ref{fig:Diff_distances_FSPL}.
The reason for this is that the FSPL of the indirect link is determined
by the \emph{product} of the distances $d_{1}$ and $d_{2}$ rather
than by their sum. Therefore, finding the optimal position for the
RIS is not a straightforward task.}
\begin{figure}[t]
\centering{}\textcolor{black}{}\subfloat[$d_{\mathrm{ris}}=40\,\mathrm{m}$.\label{fig:Rate_dif_dist_BS}]{\centering{}\textcolor{black}{\includegraphics[scale=0.8]{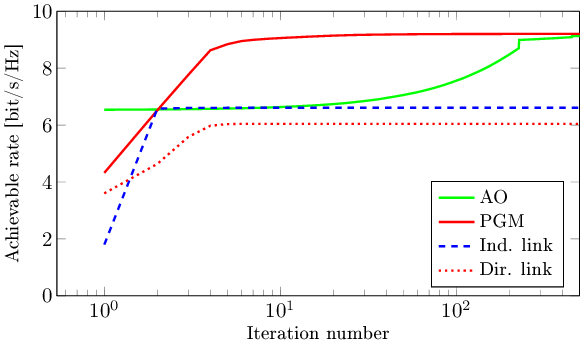}}}\textcolor{black}{\hfill{}}\subfloat[$d_{\mathrm{ris}}=D-40\,\mathrm{m}$.\label{fig:Rate_diff_dist_RX}]{\centering{}\textcolor{black}{\includegraphics[scale=0.8]{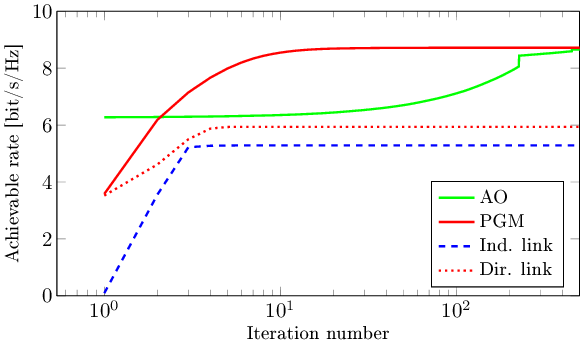}}}\textcolor{black}{\caption{\textcolor{black}{Average achievable} rate of the PGM versus the benchmark
schemes. Here $l_{t}=20\,\mathrm{m}$ and $l_{r}=100\,\mathrm{m}$.\label{fig:Rate_diff_dist}}
}
\end{figure}

\textcolor{black}{Based on the previous observations, we assume in
further simulations that the RIS is placed in the vicinity of the
transmitter ($d_{\mathrm{ris}}=40\,\mathrm{m}$) or in the vicinity
of the receiver ($d_{\mathrm{ris}}=D-40\,\mathrm{m}$). The achievable
rate results for the proposed PGM approach and for the benchmark schemes
are shown in Fig. \ref{fig:Rate_diff_dist}. It can be seen that the
proposed gradient-based optimization method converges relatively fast
to the optimum achievable rate value. On the other hand, the AO requires
significantly more iterations (at least one outer iteration, which
consists of }a sequence of\textcolor{black}{{} $N_{\mathrm{ris}}+1$
conventional iterations \cite{zhang2019capacity}) to reach its optimum
value. It can be also observed that the initial achievable rate for
the AO is higher than for the PGM. The reason for this is that for
the AO, we select from a large set of randomly generated RIS phase
shift realizations and optimized $\text{\ensuremath{\mathbf{Q}}}$
matrices the ones that provide the highest achievable rate and use
these as a starting point for the AO (thus, this specifies the initial
achievable rate). In contrast to this, the initial achievable rate
for the PGM is obtained for the aforementioned initial RIS phase shifts
and $\text{\ensuremath{\mathbf{Q}}}$ matrix.}

\textcolor{black}{In addition, the RIS is capable of providing a significant
enhancement of the achievable rate, which is proportional to the achievable
rate of the indirect link. As expected, this gain is higher when the
RIS is located in the vicinity of the transmitter, due to the lower
FSPL of the indirect link. Finally, we }observe\textcolor{black}{{}
that the FPSL ratio $T$, which is derived for a SISO system, is not
entirely trustworthy for predicting the achievable rate in a MIMO
system. Although the total FSPL of the indirect link is lower than
the total FSPL of the direct link when the RIS is placed in the vicinity
of the receiver in Fig. \ref{fig:Diff_distances_FSPL}, the direct
link will }ultimately\textcolor{black}{{} provide a higher achievable
rate in Fig.~\ref{fig:Rate_diff_dist_RX}.}
\begin{figure}[t]
\centering{}\textcolor{black}{\includegraphics[width=7cm]{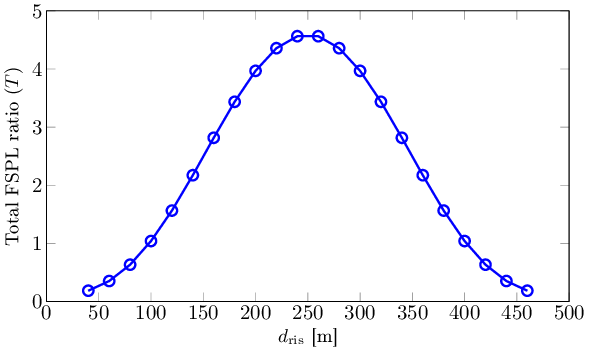}\caption{The total FSPL ratio ($T$) for the case $l_{t}=l_{r}=50\,\mathrm{m}$.\label{fig:Same_dist}}
}
\end{figure}
\textcolor{black}{}
\begin{figure}[t]
\centering{}\textcolor{black}{}\subfloat[$d_{\mathrm{ris}}=40\,\mathrm{m}$.]{\centering{}\textcolor{black}{\includegraphics[scale=0.8]{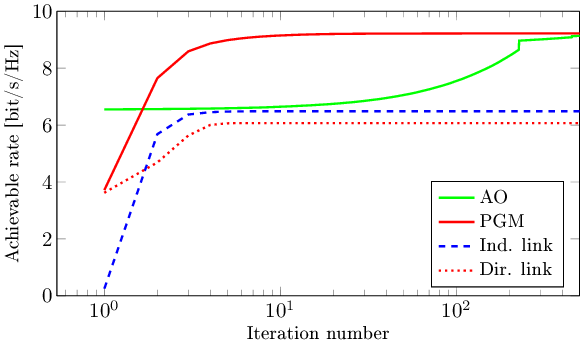}}}\textcolor{black}{\hfill{}}\subfloat[$d_{\mathrm{ris}}=D-40\,\mathrm{m}$.]{\centering{}\textcolor{black}{\includegraphics[scale=0.8]{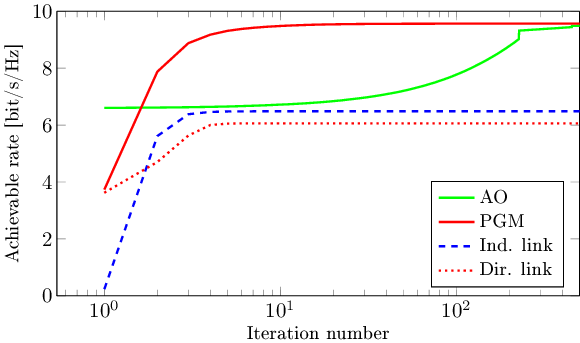}}}\textcolor{black}{\caption{\textcolor{black}{Average achievable rate of the PGM versus the benchmark
schemes. Here $l_{t}=l_{r}=50\thinspace\mathrm{m}$.\label{fig:Rate_same_dist}}}
}
\end{figure}

\textcolor{black}{In the second case, we assume $l_{t}=50\,\mathrm{m}$
and $l_{r}=50\,\mathrm{m}$. The variation of the total FSPL ratio
$T$ with the RIS position is shown in Fig. \ref{fig:Same_dist}.
It can be seen that $T$ is perfectly symmetric due to the equal values
of the distances $l_{t}$ and $l_{r}$. The same is true for the total
indirect link length, which is not shown for brevity reasons. The
achievable rate of the proposed PGM approach versus the benchmark
schemes is shown in Fig. \ref{fig:Rate_same_dist}. As expected, the
PGM has a much higher convergence rate than the AO. Also, it can be
seen that the achievable rate is slightly higher when the RIS is located
in the vicinity of the receiver than in the vicinity of the transmitter.
If the communication link are used individually, the indirect link
has a slightly higher achievable rate} than the direct link\textcolor{black}{.}

\subsubsection{\textcolor{black}{Direct link blocked}}

\begin{figure}[t]
\begin{centering}
\textcolor{black}{}\subfloat[$d_{\mathrm{ris}}=40\,\mathrm{m}$.]{\centering{}\textcolor{black}{\includegraphics[scale=0.8]{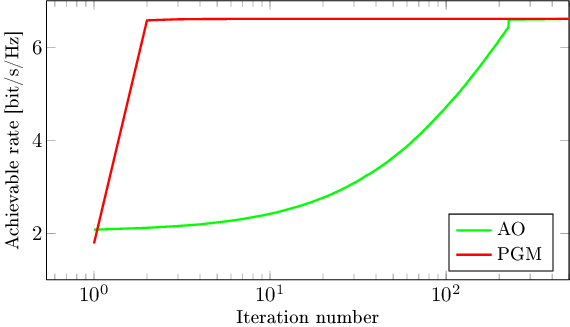}}}
\par\end{centering}
\begin{centering}
\textcolor{black}{}\subfloat[$d_{\mathrm{ris}}=D-40\,\mathrm{m}$.\label{fig:Rate_no_DIR}]{\centering{}\textcolor{black}{\includegraphics[scale=0.8]{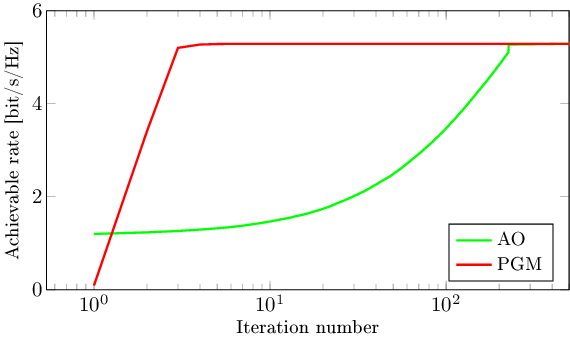}}}
\par\end{centering}
\centering{}\textcolor{black}{\caption{\textcolor{black}{Average achievable rate of the PGM versus the AO.
The parameter setup is the same as in Fig. \ref{fig:Diff_distances}.\label{fig:Rate-blocked}}}
}
\end{figure}

\textcolor{black}{If the direct link between the transmitter and the
receiver is blocked, the only means of signal transmission is via
the RIS. It can be easily seen that the main observations made concerning
the optimal RIS position in the previous subsection are also applicable
here. Therefore, we analyze the achievable rate when the RIS is placed
in the vicinity of the transmitter or in the vicinity of the receiver.
The achievable rate results of the PGM versus the AO are shown in
Fig. \ref{fig:Rate-blocked}. }In both cases, the\textcolor{black}{{}
optimal achievable}\footnote{\textcolor{black}{In our simulations, we take the ``optimal achievable
rate'' to be that which is obtained at the final (i.e., 500th) iteration.}}\textcolor{black}{{} rates match the achievable rates of the second
benchmark scheme in Fig. \ref{fig:Diff_distances}. The PGM requires
only a few iterations to converge to the optimum value. }On the other
hand\textcolor{black}{, the AO needs approximately $N_{\mathrm{ris}}+1$
iterations (i.e., one outer iteration) to reach the optimum value.
Interestingly, the achievable rate enhancement during the first outer
iteration is higher when the direct link is blocked. It seems that
the absence of the direct link }may have\textcolor{black}{{} a significant
influence on choosing the initial covariance matrix and RIS phase
shifts, before starting the AO.}

\subsection{\textcolor{black}{Scaling with $N_{\mathrm{ris}}$}}

\textcolor{black}{This subsection consists of three parts. First,
we demonstrate the correctness of the expression \eqref{eq:pow_equ}
in Section \ref{sec:Total-FSPL}. Then, we show how increasing the
number of RIS elements influences the achievable rate of the considered
system. Finally, we present the trade-off between the operating frequency
and the number of RIS elements. }
\begin{figure}[t]
\begin{centering}
\textcolor{black}{\includegraphics[scale=0.8]{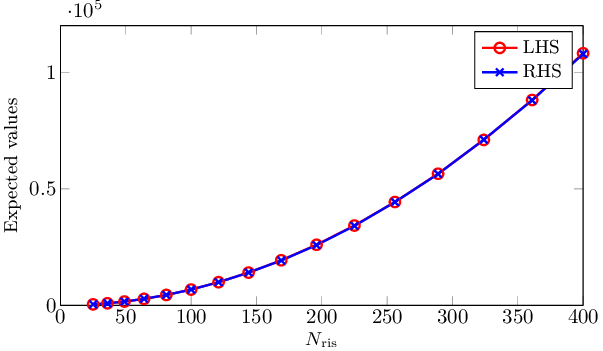}}
\par\end{centering}
\textcolor{black}{\caption{Comparison of the left-hand side and right-hand side of \eqref{eq:pow_equ}.\label{fig:Side_comp}}
}
\end{figure}
\textcolor{black}{}
\begin{figure}[tbh]
\begin{centering}
\textcolor{black}{\includegraphics[scale=0.8]{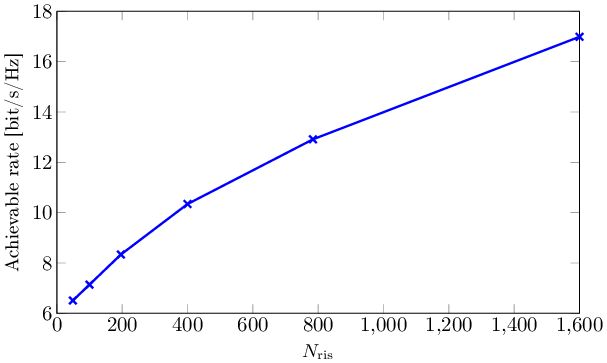}}
\par\end{centering}
\centering{}\textcolor{black}{\caption{\textcolor{black}{Average achiev}able rate versus the number of RIS
elements $N_{\mathrm{ris}}$.\label{fig:RateVsNris}}
}
\end{figure}

\textcolor{black}{To verify the correctness of the upper-bound expression
in \eqref{eq:pow_equ}, we compare the values of the left and right
hand sides of this expression in Fig. \ref{fig:Side_comp}. As the
expression pertains to single-antenna systems, we assume $N_{t}=N_{r}=1$
in this simulation. For completeness, the presented results are computed
and averaged for different positions of the RIS. The graph shows a
very good match between the two sides of the aforementioned expression,
which means that in practice the total FSPL of the indirect link in
a SISO system is very well approximated by \eqref{eq:Total_FPSL_indir}.}

\textcolor{black}{In general, it is not easy to assess the expected
achievable rate for some arbitrary value of $N_{\mathrm{ris}}$, when
gradient-based optimization methods are applied. Therefore, to obtain
a better understanding of the variation of the achievable rate with
$N_{\mathrm{ris}}$, we present a numerical evaluation of the achievable
rate in Fig.~\ref{fig:RateVsNris}. The parameter setup is the same
as for Fig.~\ref{fig:Diff_distances} and $d_{\mathrm{ris}}=40\,\mathrm{m}$.
In this case, the physical size of the RIS is actually increasing,
while the RIS is always operating in the far field \cite{danufane2020path}.
As a result, we observe that there is an increase in the achievable
rate when $N_{\mathrm{ris}}$ is doubled}\footnote{\textcolor{black}{It should be noted that the number of RIS elements
is not exactly doubled in simulations, since we aim to have a square
RIS. Therefore, the achievable rate is computed and plotted for $N_{\mathrm{ris}}$
equal to 49, 196 and 784 instead of 50, 200 and 800, respectively.}}\textcolor{black}{{} and this increase becomes larger as we increase
$N_{\mathrm{ris}}$. Also, the slope of the achievable rate curve
gradually reduces with $N_{\mathrm{ris}}$, as a consequence of the
logarithm function in the achievable rate expression.}

\textcolor{black}{}
\begin{figure}[t]
\centering{}\textcolor{black}{\includegraphics[scale=0.75]{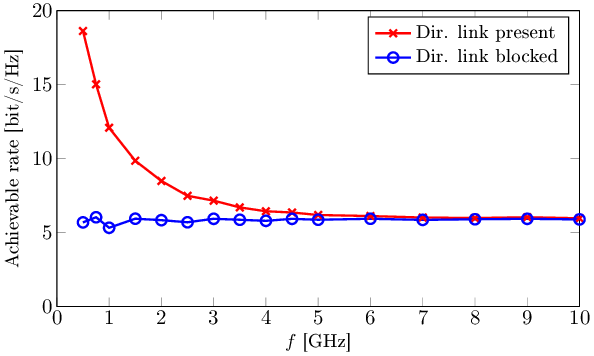}\caption{\textcolor{black}{Average achievable rate versus frequency $f$. The
parameter setup is the same as for Fig. \ref{fig:Diff_distances}.\label{fig:Achievable-rate_vs_freq}}}
}
\end{figure}
\textcolor{black}{The number of RIS elements that can be placed on
an RIS having a constant physical size, without causing coupling between
the neighboring RIS elements, increases with the frequency of operation.
}Therefore, the RIS may consist of a large number of RIS elements,
if it is intended to work at high frequencies.\textcolor{black}{{} Motivated
by this fact, we analyze the trade-off between the operating frequency
and the number of RIS elements $N_{\mathrm{ris}}$, and their influence
on the achievable rate in the considered system. We assume that the
RIS elements are placed in an RIS of size $1\,\mathrm{m}\times1\,\mathrm{m}$
and $s_{\mathrm{ris}}=\lambda/2$ at all frequencies. The achievable
rate of the PGM versus the operating frequency $f$ is shown in Fig.
\ref{fig:Achievable-rate_vs_freq}. In the frequency range up to $5\,\mathrm{GHz}$,
the achievable rate of the considered system decreases primarily because
of the FSPL increase of the direct link. At higher frequencies, the
achievable rate remains almost constant regardless of whether the
direct link is present or blocked. In other words, the FSPL of the
direct link is so high in this case that the direct link becomes practically
useless for communicating information. On the other hand, the indirect
link has approximately the same achievable rate across the entire
frequency range. It is because the increase in the FSPL of the indirect
link is compensated by the increased number of RIS elements in the
considered system. Finally, we conclude that the direct link is only
useful at lower frequencies, while the indirect link can be used at
all frequencies if a sufficient number of RIS elements is provided.}

\subsection{\textcolor{black}{Achievable Rate in Indoor Environments}}

\textcolor{black}{All of the previous simulation results are obtained
for a wireless communication system operating in an outdoor environment.
To further demonstrate the effectiveness of the proposed gradient-based
optimization method, we consider its implementation in an indoor environment.
Since the communication distances are now much smaller and the communication
bandwidths are usually larger (i.e., typically 20/22~MHz), the following
simulation parameters have the following altered values: $D=30\,\mathrm{m}$,
$d_{\mathrm{ris}}=5\,\mathrm{m}$, $l_{t}=3\,\mathrm{m}$, $l_{r}=7\,\mathrm{m}$,
$N_{\mathrm{ris}}=100$, $P_{t}=-30\,\mathrm{dB}$ and $N_{0}=-100\thinspace\mathrm{dB}$.}

\textcolor{black}{}
\begin{figure}[t]
\begin{centering}
\textcolor{black}{}\subfloat[Direct link present.]{\centering{}\textcolor{black}{\includegraphics[scale=0.8]{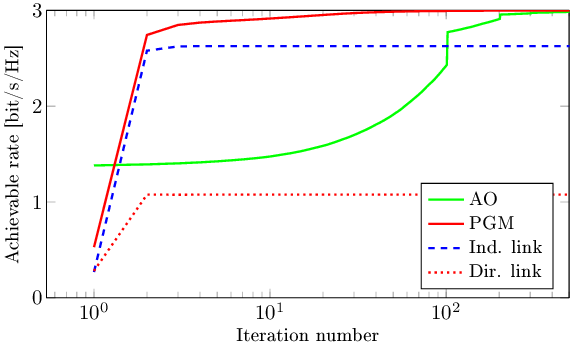}}}
\par\end{centering}
\begin{centering}
\textcolor{black}{}\subfloat[Direct link blocked.]{\centering{}\textcolor{black}{\includegraphics[scale=0.8]{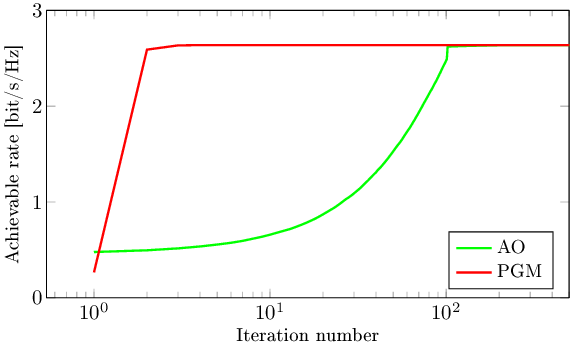}}}
\par\end{centering}
\textcolor{black}{\caption{\textcolor{black}{Average achievable} rate of the PGM versus the benchmark
schemes in an indoor environment.\label{fig:Rate_indoor}}
}
\end{figure}
\textcolor{black}{The simulation results of the proposed optimization
method versus the benchmark schemes in an indoor environment are presented
in Fig. \ref{fig:Rate_indoor}. The PGM again requires a lower number
of iterations than the AO to converge to the optimal achievable rate.
In contrast to the previous simulation results, the achievable rate
in indoor environments is almost entirely determined by the indirect
link signal transmission, which can explained by the following argument.
Reducing the distances in the considered communication system results
in the reduction of the total \acp{FSPL}, and the total \ac{FSPL}
of the indirect link is particularly affected by this, since it is
inversely proportional to the product of distances. Hence, a very
small number of RIS elements is sufficient to enable the indirect
link to have a lower total \ac{FSPL} than the direct link, and any
further increase of the number of RIS elements will render the direct
link comparatively useless for indoor communications.}

\subsection{\textcolor{black}{Computational Complexity and Run Time Results\label{subsec:Comp-compl-results}}}

\textcolor{black}{In reality, it is not practical to the wait for
an optimization algorithm to reach a critical point, but rather some
value that is not too far from it. Hence} in this subsection\textcolor{black}{,
we consider the computational complexity required for the PGM and
the AO to reach an achievable rate that is equal to 95\,\% of the
average achievable rate at the 500th iteration. These complexities
are heavily influenced by the number of iterations that are needed
to achieve this target achievable rate. For the \ac{PGM}, the computational
complexity per iteration is $C_{\mathrm{PGM,IT}}$ given by \eqref{eq:pgm_comp}
and the number of iterations needed to achieve the optimal achievable
rate is denoted as $I_{\mathrm{PGM}}$. Their product determines the
total computational complexity}\footnote{\textcolor{black}{We neglect the number of multiplications needed
to compute the achievable rate after every iteration, due to the low
number of iterations that PGM needs to reach the target achievable
rate.}}\textcolor{black}{{} $C_{\mathrm{PGM}}$ of the \ac{PGM}. For the \ac{AO},
the computational complexity is given by}\footnote{\textcolor{black}{This complexity is derived under the assumption
that the achievable rate is computed at the end of each outer iteration,
as was proposed in \cite{zhang2019capacity}. Therefore, we are not
interested in the number of conventional iterations, but in the number
of \emph{outer} iterations needed to achieve a certain rate.}}\textcolor{black}{{} $C_{\mathrm{AO}}$ and the number of outer iterations
needed to achieve the optimal achievable rate is $I_{\mathrm{OI}}$
(see Appendix \ref{sec:AO_comp}). To maintain compatibility with
\cite{zhang2019capacity}, the number of randomly generated RIS phase
shift realizations at the beginning of the \ac{AO} is taken to be
$L_{\mathrm{AO}}=100$. }

\textcolor{black}{The computational complexity of the \ac{PGM} and
of the \ac{AO} is shown in Table \ref{tab:Comp}. The parameter setup
is the same as for Figs. \ref{fig:Rate_diff_dist_RX} and \ref{fig:Rate_no_DIR}.
In general, the PGM is able to achieve a significantly lower computational
complexity than the \ac{AO}, while at the same time it requires a
small number of iterations. If the direct link is present, we observe
that $I_{\mathrm{PGM}}$ becomes smaller with an increase of the number
of RIS elements, or in other words, the convergence of the \ac{PGM}
improves. As a result, the computational complexity of the \ac{PGM}
does not increase proportionally to the number of RIS elements. On
the other hand, $I_{\mathrm{PGM}}$ remains constant when the direct
link is blocked and the computational complexity of the \ac{PGM}
increases if the number of RIS elements is made larger. The AO needs
one outer iteration to reach the target achievable rate, and its computational
complexity increases in proportion to the number of RIS elements.}

\begin{figure}[t]
\centering{}\textcolor{black}{\includegraphics[scale=0.75]{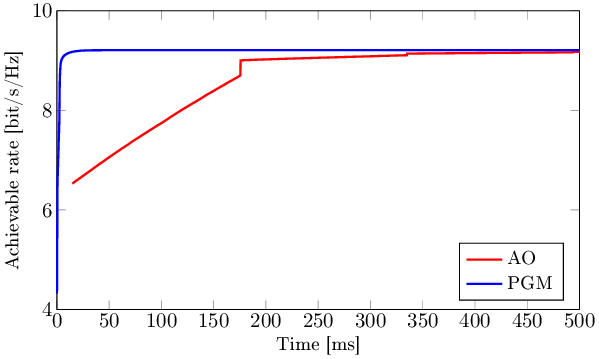}\caption{Average achievable rate of the PGM and the AO versus the run time.
The parameter setup is the same as for Fig. \ref{fig:Rate_dif_dist_BS}.\textcolor{blue}{{}
\label{fig:Rate_vs_time}}}
}
\end{figure}

\textcolor{black}{To make this subsection complete, we also compare
in Fig.~\ref{fig:Rate_vs_time} the achievable rate of the AO and
the PGM with respect to the run time of the algorithm's software implementation.
The achievable rate for both methods is computed at the end of each
iteration. It can be seen that the PGM needs an extremely low run
time to converge. Approximately the same time is needed for the AO
just to select the optimal initial point. Since the first iteration
of the AO is executed after the initial point is chosen, the achievable
rate curve for the AO in Fig. \ref{fig:Rate_vs_time} starts from
about 10\,ms. Even after 500\,ms the AO is not entirely capable
of reaching the same achievable rate.}

\subsection{\textcolor{black}{Sensitivity of PGM to Initialization}}

\begin{figure}[t]
\centering{}\textcolor{black}{\includegraphics[scale=0.75]{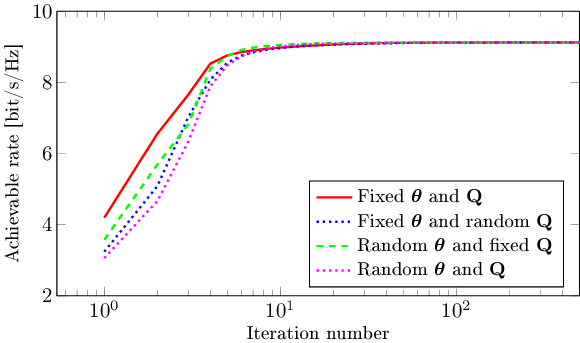}\caption{\textcolor{black}{Average achievable rate results for different initial
values of $\boldsymbol{\theta}$ and $\mathbf{Q}$. The parameter
setup is the same as for Fig. \ref{fig:Rate_dif_dist_BS}}. \label{fig:Sensitivity}}
}
\end{figure}

\textcolor{black}{In this subsection, we study the sensitivity of
the PGM to the initial values of $\boldsymbol{\theta}$ and $\mathbf{Q}$.
Hence, we consider four cases, where the initial value of $\boldsymbol{\theta}$
is either set to $[1\;1\;\cdots\;1]^{T}$ (referred to as ``fixed
$\boldsymbol{\theta}$'') or is randomly generated, and the initial
value of $\mathbf{Q}$ is either set to $(P_{t}/N_{t})\mathbf{I}$
(referred to as ``fixed $\mathbf{Q}$'') or is randomly generated.
The achievable rate results for the different initial values of $\boldsymbol{\theta}$
and $\mathbf{Q}$ are shown in Fig. \ref{fig:Sensitivity}. The only
visible difference between the considered cases is in the first few
iterations, where the PGM with fixed initial $\boldsymbol{\theta}$
and $\mathbf{Q}$ achieves a slightly higher achievable rate than
in other cases. In later iterations, the achievable rates in all four
cases are approximately equal. Hence, the PGM can always reach the
same achievable rate in approximately the same number of iterations,
independently of the initial values of $\boldsymbol{\theta}$ and
$\mathbf{Q}$.}

\subsection{\textcolor{black}{Robustness to System Imperfections}}

\begin{figure}[t]
\centering{}\textcolor{black}{\includegraphics[scale=0.75]{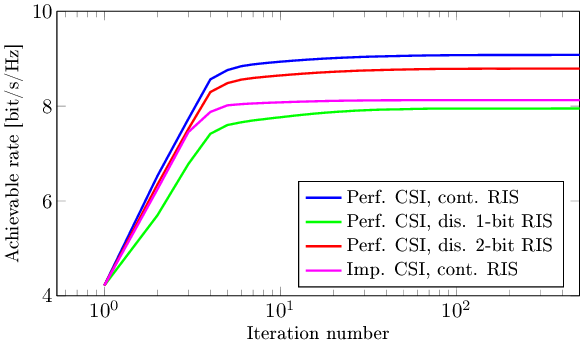}\caption{\textcolor{black}{Average achievable rate for the case of discrete
RIS phase shifts and imperfect CSI. The parameter setup is the same
as for Fig. \ref{fig:Rate_dif_dist_BS}.}\textcolor{blue}{\label{fig:Rate-DisPhShift-CSI}}}
}
\end{figure}

\textcolor{black}{In order to better understand the applicability
of the proposed PGM, it is necessary to consider the influence of
realistic imperfections in an RIS-aided communication system. Motivated
by this, the achievable rate for the case of discrete RIS phase shifts
and imperfect \ac{CSI} is shown in Fig. \ref{fig:Rate-DisPhShift-CSI}.
The achievable rate for the RIS with discrete phase shifts is obtained
by discretizing the continuous RIS phase shifts in the final iteration
of the PGM and then calculating the achievable rate. The proposed
PGM is also directly applicable to discrete phase shifts, since the
projection of a given point onto the set of discrete phase shifts
is equivalent to finding the minimum distance between the point and
all possible phase shifts. It can be seen that utilizing 1-bit and
2-bit discrete RIS phase shifts can reduce the optimal achievable
rate by approximately 1.1\,bit/s/Hz and 0.2\,bit/s/Hz, respectively.
Hence, even a very low resolution of discrete RIS phase shifts is
sufficient to ensure a limited reduction of the optimal achievable
rate.}
\textcolor{black}{{} }

\textcolor{black}{In the case of imperfect \ac{CSI}, we assume that
the estimated channel matrix can be presented as a sum of the true
channel matrix and an estimation error matrix. The estimation error
matrix consists of \ac{iid} elements that are distributed according
to $\mathcal{CN}(0,\sigma^{2})$, where $\sigma^{2}=0.2$. Also, it
is assumed that the channel matrix \acp{FSPL} are not affected by
imperfect \ac{CSI}. From the results, which are plotted in Fig. \ref{fig:Rate-DisPhShift-CSI},
we can observe that the optimal achievable rate decreases by approximately
1\,bit/s/Hz, which is an acceptable level of reduction.}

\subsection{\textcolor{black}{Influence of Data Scaling and Line Search}}

\begin{figure}[t]
\centering{}\textcolor{black}{\includegraphics[scale=0.75]{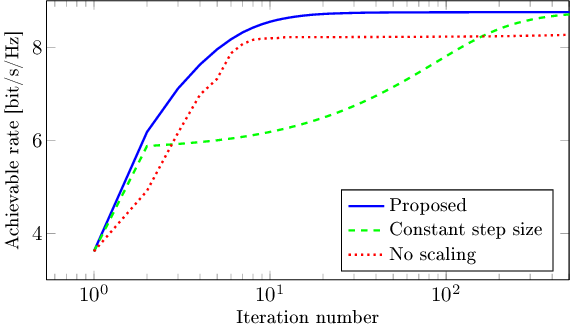}\caption{Average achievable rate of the proposed PGM and the two benchmark
schemes (i.e., PGM without line search and PGM without data scaling).
The setup of parameters is the same as for Fig. \ref{fig:Rate_diff_dist_RX}.\label{fig:PGM_all}}
}
\end{figure}

\textcolor{black}{In this subsection, we analyze the influence of
data scaling and line search on the PGM. Hence, we compare the proposed
PGM with two benchmark schemes. For the first benchmark scheme (i.e.,
PGM without line search), the PGM is implemented without the line
search procedure and we assumed a constant step size equal to 10.
For the second benchmark scheme, the PGM is implemented without data
scaling. The achievable rate results are shown in Fig. \ref{fig:PGM_all}.
As expected, the proposed PGM has the best achievable rate results
among the considered schemes. PGM without line search needs significantly
more iterations to reach the optimal achievable rate, for a step size
that is multiple times larger than the inverse of the Lipschitz constant
(see Theorem \ref{thm:convergence}). Generally, the larger step size
enables faster convergence, but the risk of misconvergence is then
higher. Furthermore, PGM without data scaling has an achievable rate
that is not very significantly worse than the achievable rate of the
proposed PGM.}

\section{\textcolor{black}{Conclusion\label{sec:Conclusion}}}

\textcolor{black}{In this paper, we proposed a new \ac{PGM} algorithm
for the achievable rate optimization in multi-stream MIMO system equipped
with an RIS. Also, we derived a Lipschitz constant that guarantees
the convergence of the \ac{PGM}. To improve the rate of convergence
of the \ac{PGM} algorithm, we proposed a data scaling step and employed
a backtracking line search, which enable the \ac{PGM} to significantly
outperform the existing AO algorithm. In addition, we defined the
new metric of total FSPL, and showed that the ratio between the total
FSPL of the indirect and direct links can successfully serve as a
first-order measure of the applicability of an RIS. Numerical results
confirm that the PGM requires a significantly lower number of iterations,
and correspondingly a substantially lower computational complexity,
than the AO in order to reach a target (near-optimal) achievable rate.
Furthermore, we showed that the RIS is particularly convenient for
application in an indoor environment, since a small number of RIS
elements is sufficient to enable the indirect link to have a higher
achievable rate than the direct link.  }

\appendices{}

\section{\textcolor{black}{Complex-valued Gradient of $f(\boldsymbol{\theta},\mathbf{Q})$\label{sec:grad}}}

\textcolor{black}{We first note that \eqref{eq:grad_Q} is given in
\cite[Eq. (6.207)]{Are2011} and is relatively well known in the related
literature. To derive \eqref{eq:grad_theta} we follow the procedure
to compute the complex-valued gradient of a general function detailed
in \cite[Sect. 3.3.1]{Are2011}. Note that in the following, we adopt
the notations introduced in \cite{Are2011}: $df(\mathbf{X})$ denotes
the complex differential of $f(\mathbf{X})$. To proceed, we recall
that the complex differential of $f(\boldsymbol{\theta},\mathbf{Q})$
with respect to $\mathbf{F}(\boldsymbol{\theta})=\diag(\boldsymbol{\theta})$
and $\mathbf{F}^{*}(\boldsymbol{\theta})$ is given by
\begin{gather}
df(\boldsymbol{\theta},\mathbf{Q})=\tr\left\{ \mathbf{K}(\boldsymbol{\theta},\mathbf{Q})d\left(\mathbf{Z(\boldsymbol{\theta})}\mathbf{Q}\mathbf{Z}^{H}(\boldsymbol{\theta})\right)\right\} =\nonumber \\
\tr\left\{ \mathbf{K}(\boldsymbol{\theta},\mathbf{Q})\left(d(\mathbf{Z(\boldsymbol{\theta})})\mathbf{Q}\mathbf{Z}^{H}(\boldsymbol{\theta})+\mathbf{Z(\boldsymbol{\theta})}\mathbf{Q}d\mathbf{Z}^{H}(\boldsymbol{\theta})\right)\right\} .
\end{gather}
After a few algebraic steps we obtain
\begin{multline}
df(\boldsymbol{\theta},\mathbf{Q})=\vect^{T}\left(\left(\bar{\mathbf{H}}_{1}\mathbf{Q}\mathbf{Z}^{H}(\boldsymbol{\theta})\mathbf{K}(\boldsymbol{\theta},\mathbf{Q})\mathbf{H}_{2}\right)^{T}\right)\vect(d\mathbf{F}(\boldsymbol{\theta}))\\
+\vect^{T}\left(\left(\bar{\mathbf{H}}_{1}^{\ast}\mathbf{Z}^{T}(\boldsymbol{\theta})\mathbf{Q}^{T}\mathbf{K}(\boldsymbol{\theta},\mathbf{Q})\mathbf{H}_{2}^{\ast}\right)^{T}\right)\vect\left(d\mathbf{F}^{*}(\boldsymbol{\theta})\right),
\end{multline}
where we have used the equality $\tr(\mathbf{A}^{T}\mathbf{B})=\vect^{T}(\mathbf{A})\vect(\mathbf{B})$.
Let $\mathbf{L}_{d}$ be the matrix used to place the diagonal elements
of a square matrix $\mathbf{A}$ on $\vect(\mathbf{A})$, i.e. $\vect(\mathbf{A})=\mathbf{L}_{d}\vect_{d}(\mathbf{A})$
\cite[Definition 2.12]{Are2011}. Then we can rewrite $df(\boldsymbol{\theta},\mathbf{Q})$
as
\begin{multline}
df(\boldsymbol{\theta},\mathbf{Q})=\vect^{T}\left(\left(\bar{\mathbf{H}}_{1}\mathbf{Q}\mathbf{Z}^{H}(\boldsymbol{\theta})\mathbf{K}(\boldsymbol{\theta},\mathbf{Q})\mathbf{H}_{2}\right)^{T}\right)\mathbf{L}_{d}\vect(d\boldsymbol{\theta})\\
+\vect^{T}\left(\left(\bar{\mathbf{H}}_{1}^{\ast}\mathbf{Z}^{T}(\boldsymbol{\theta})\mathbf{Q}^{T}\mathbf{K}(\boldsymbol{\theta},\mathbf{Q})\mathbf{H}_{2}^{\ast}\right)^{T}\right)\mathbf{L}_{d}\vect(d\boldsymbol{\theta}^{\ast}).
\end{multline}
Using \cite[Table 3.2]{Are2011} and \cite[Eqn. (2.140)]{Are2011}
we obtain
\begin{alignat}{1}
\nabla_{\theta}f(\boldsymbol{\theta},\mathbf{Q}) & =\mathbf{L}_{d}^{T}\vect\left(\mathbf{H}_{2}^{H}\mathbf{K}(\boldsymbol{\theta},\mathbf{Q})\mathbf{Z}(\boldsymbol{\theta})\mathbf{Q}\bar{\mathbf{H}}_{1}^{H}\right)\nonumber \\
 & =\vect_{d}\left(\mathbf{H}_{2}^{H}\mathbf{K}(\boldsymbol{\theta},\mathbf{Q})\mathbf{Z}(\boldsymbol{\theta})\mathbf{Q}\bar{\mathbf{H}}_{1}^{H}\right).
\end{alignat}
In a similar manner, we can prove the expression for $\nabla_{\mathbf{Q}}f(\boldsymbol{\theta},\mathbf{Q})$.
The details are omitted here due to the page limit.}

\section{\textcolor{black}{Proof of Lemma \ref{lem:Lipschitz:part}\label{sec:Proof:lipschitz:part}}}

\textcolor{black}{}%
\begin{comment}
\textcolor{black}{To make the proof accessible to a reader first we
need to introduce some inequalities (without proof) that we use in
this section. Then we can refer to these inequalities}
\end{comment}
\textcolor{black}{To make the proof easy to follow, we first recall
the following inequalities, which are well known or can be proved
easily. For the norm of a matrix product it holds that\begin{subequations}
\begin{align}
||\mathbf{A}\mathbf{B}|| & \leq\lambda_{\max}(\mathbf{A})||\mathbf{B}||\\
||\mathbf{A}\mathbf{B}\mathbf{C}|| & \leq\lambda_{\max}(\mathbf{A})||\mathbf{B}||\lambda_{\max}(\mathbf{C})
\end{align}
\end{subequations}where $\lambda_{\max}(\mathbf{X})$ denotes the
largest singular value of $\lambda_{\max}(\mathbf{X})$. Since $\mathbf{K}(\bar{\boldsymbol{\theta}},\bar{\mathbf{Q}})=\left(\mathbf{I}+\mathbf{Z}(\bar{\boldsymbol{\theta}})\bar{\mathbf{Q}}\mathbf{Z}(\bar{\boldsymbol{\theta}})^{H}\right)^{-1}\preceq\mathbf{I}$,
we have
\begin{equation}
\lambda_{\max}\left(\mathbf{K}(\bar{\boldsymbol{\theta}},\bar{\mathbf{Q}})\right)\leq1.
\end{equation}
It is easy to check that
\begin{equation}
\lambda_{\max}\left(\mathbf{F}(\bar{\boldsymbol{\theta}})\right)=k^{-1};\lambda_{\max}\left(\bar{\mathbf{Q}}\right)\le\bar{P}_{t}.
\end{equation}
}

\subsection{\textcolor{black}{Proof of \eqref{eq:grad_phi_dif_fin}}}

\textcolor{black}{From \eqref{eq:grad_theta} we obtain
\begin{multline}
||\nabla_{\mathbf{\bar{\boldsymbol{\theta}}}}f(\bar{\boldsymbol{\theta}}_{1},\bar{\mathbf{Q}}_{1})-\nabla_{\mathbf{\bar{\boldsymbol{\theta}}}}f(\bar{\boldsymbol{\theta}}_{2},\bar{\mathbf{Q}}_{2})||=||\mathbf{H}_{2}^{H}\mathbf{K}(\bar{\boldsymbol{\theta}}_{1},\bar{\mathbf{Q}}_{1})\mathbf{Z}(\bar{\boldsymbol{\theta}}_{1})\bar{\mathbf{Q}}_{1}\mathbf{\bar{\mathbf{H}}}_{1}^{H}\\
-\mathbf{H}_{2}^{H}\mathbf{K}(\bar{\boldsymbol{\theta}}_{2},\bar{\mathbf{Q}}_{2})\mathbf{Z}(\bar{\boldsymbol{\theta}}_{2})\bar{\mathbf{Q}}_{2}\mathbf{\bar{\mathbf{H}}}_{1}^{H}||\le||\mathbf{H}_{2}^{H}\mathbf{K}(\bar{\boldsymbol{\theta}}_{1},\bar{\mathbf{Q}}_{1})\mathbf{Z}(\bar{\boldsymbol{\theta}}_{1})\bar{\mathbf{Q}}_{1}\mathbf{\bar{\mathbf{H}}}_{1}^{H}\\
-\mathbf{H}_{2}^{H}\mathbf{K}(\bar{\boldsymbol{\theta}}_{1},\bar{\mathbf{Q}}_{1})\mathbf{Z}(\bar{\boldsymbol{\theta}}_{2})\bar{\mathbf{Q}}_{2}\mathbf{\bar{\mathbf{H}}}_{1}^{H}||+||\mathbf{H}_{2}^{H}\mathbf{K}(\bar{\boldsymbol{\theta}}_{1},\bar{\mathbf{Q}}_{1})\mathbf{Z}(\bar{\boldsymbol{\theta}}_{2})\bar{\mathbf{Q}}_{2}\mathbf{\bar{\mathbf{H}}}_{1}^{H}\\
-\mathbf{H}_{2}^{H}\mathbf{K}(\bar{\boldsymbol{\theta}}_{2},\bar{\mathbf{Q}}_{2})\mathbf{Z}(\bar{\boldsymbol{\theta}}_{2})\bar{\mathbf{Q}}_{2}\bar{\mathbf{H}}_{1}^{H}||.\label{eq:grad_phi_dif}
\end{multline}
The first term on the right-hand side of \eqref{eq:grad_phi_dif}
can be upper-bounded as
\begin{multline}
\!\!\!\!\!\!||\mathbf{H}_{2}^{H}\mathbf{K}(\bar{\boldsymbol{\theta}}_{1},\bar{\mathbf{Q}}_{1})\mathbf{Z}(\bar{\boldsymbol{\theta}}_{1})\bar{\mathbf{Q}}_{1}\mathbf{\bar{\mathbf{H}}}_{1}^{H}-\mathbf{H}_{2}^{H}\mathbf{K}(\bar{\boldsymbol{\theta}}_{1},\bar{\mathbf{Q}}_{1})\mathbf{Z}(\bar{\boldsymbol{\theta}}_{2})\bar{\mathbf{Q}}_{2}\mathbf{\bar{\mathbf{H}}}_{1}^{H}||\\
\le a\lambda_{\max}(\mathbf{\bar{\mathbf{H}}}_{\mathrm{DIR}})||\bar{\mathbf{Q}}_{1}-\bar{\mathbf{Q}}_{2}||\\
+a\lambda_{\max}(\mathbf{H}_{2})||\mathbf{F}(\bar{\boldsymbol{\theta}}_{1})\bar{\mathbf{H}}_{1}\bar{\mathbf{Q}}_{1}-\mathbf{F}(\bar{\boldsymbol{\theta}}_{2})\mathbf{\bar{\mathbf{H}}}_{1}\bar{\mathbf{Q}}_{2}||.\label{eq:phi_first_part}
\end{multline}
Furthermore, we have
\begin{multline}
||\mathbf{F}(\bar{\boldsymbol{\theta}}_{1})\mathbf{\bar{\mathbf{H}}}_{1}\bar{\mathbf{Q}}_{1}-\mathbf{F}(\bar{\boldsymbol{\theta}}_{2})\mathbf{\bar{\mathbf{H}}}_{1}\bar{\mathbf{Q}}_{2}||\le k^{-1}\lambda_{\max}(\mathbf{\bar{\mathbf{H}}}_{1})||\bar{\mathbf{Q}}_{1}-\bar{\mathbf{Q}}_{2}||\\
+\lambda_{\max}(\bar{\mathbf{H}}_{1})\bar{P}_{t}||\bar{\boldsymbol{\theta}}_{1}-\bar{\boldsymbol{\theta}}_{2}||.\label{eq:phi_first_part01}
\end{multline}
Substituting \eqref{eq:phi_first_part01} into \eqref{eq:phi_first_part}
gives
\begin{multline}
\!\!\!\!\!\!||\mathbf{H}_{2}^{H}\mathbf{K}(\bar{\boldsymbol{\theta}}_{1},\bar{\mathbf{Q}}_{1})\mathbf{Z}(\bar{\boldsymbol{\theta}}_{1})\bar{\mathbf{Q}}_{1}\mathbf{\bar{\mathbf{H}}}_{1}^{H}-\mathbf{H}_{2}^{H}\mathbf{K}(\bar{\boldsymbol{\theta}}_{1},\bar{\mathbf{Q}}_{1})\mathbf{Z}(\bar{\boldsymbol{\theta}}_{2})\bar{\mathbf{Q}}_{2}\mathbf{\bar{\mathbf{H}}}_{1}^{H}||\\
\le ab||\bar{\mathbf{Q}}_{1}-\bar{\mathbf{Q}}_{2}||+a^{2}\bar{P}_{t}||\bar{\boldsymbol{\theta}}_{1}-\bar{\boldsymbol{\theta}}_{2}||.\label{eq:phi_first_part_fin}
\end{multline}
Similarly, the second term on the right-hand side (RHS) of \eqref{eq:grad_phi_dif}
can be upper-bounded as
\begin{multline}
\!\!\!\!\!\!||\mathbf{H}_{2}^{H}\mathbf{K}(\bar{\boldsymbol{\theta}}_{1},\bar{\mathbf{Q}}_{1})\mathbf{Z}(\bar{\boldsymbol{\theta}}_{2})\bar{\mathbf{Q}}_{2}\mathbf{\bar{\mathbf{H}}}_{1}^{H}-\mathbf{H}_{2}^{H}\mathbf{K}(\bar{\boldsymbol{\theta}}_{2},\bar{\mathbf{Q}}_{2})\mathbf{Z}(\bar{\boldsymbol{\theta}}_{2})\bar{\mathbf{Q}}_{2}\mathbf{\bar{\mathbf{H}}}_{1}^{H}||\\
\le ab\bar{P}_{t}||\mathbf{Z}(\bar{\boldsymbol{\theta}}_{2})\bar{\mathbf{Q}}_{2}\mathbf{Z}(\bar{\boldsymbol{\theta}}_{2})^{H}-\mathbf{Z}(\bar{\boldsymbol{\theta}}_{1})\bar{\mathbf{Q}}_{1}\mathbf{Z}(\bar{\boldsymbol{\theta}}_{1})^{H}||.\label{eq:phi_second_part}
\end{multline}
}%
\begin{comment}
\textcolor{black}{we may mention that $a$ and $b$ are defined previously
in ()}
\end{comment}
\textcolor{black}{Furthermore, we obtain
\begin{gather}
||\mathbf{Z}(\bar{\boldsymbol{\theta}}_{2})\bar{\mathbf{Q}}_{2}\mathbf{Z}(\bar{\boldsymbol{\theta}}_{2})^{H}-\mathbf{Z}(\bar{\boldsymbol{\theta}}_{1})\bar{\mathbf{Q}}_{1}\mathbf{Z}(\bar{\boldsymbol{\theta}}_{1})^{H}||\nonumber \\
\le||\mathbf{Z}(\bar{\boldsymbol{\theta}}_{2})\bar{\mathbf{Q}}_{2}(\mathbf{Z}(\bar{\boldsymbol{\theta}}_{2})^{H}-\mathbf{Z}(\bar{\boldsymbol{\theta}}_{1})^{H})||\nonumber \\
+||(\mathbf{Z}(\bar{\boldsymbol{\theta}}_{2})\bar{\mathbf{Q}}_{2}-\mathbf{Z}(\bar{\boldsymbol{\theta}}_{1})\bar{\mathbf{Q}}_{1})\mathbf{Z}(\bar{\boldsymbol{\theta}}_{1})^{H}||.\label{eq:phi_second_part_1}
\end{gather}
The following inequalities hold for the two norms in the RHS of the
above equation:
\begin{multline}
\!\!\!\!\!\!||\mathbf{Z}(\bar{\boldsymbol{\theta}}_{2})\bar{\mathbf{Q}}_{2}(\mathbf{Z}(\bar{\boldsymbol{\theta}}_{2})^{H}-\mathbf{Z}(\bar{\boldsymbol{\theta}}_{1})^{H})||=||(\mathbf{\bar{\mathbf{H}}}_{\mathrm{DIR}}+\mathbf{H}_{2}\mathbf{F}(\bar{\boldsymbol{\theta}}_{2})\mathbf{\bar{\mathbf{H}}}_{1})\\
\times\bar{\mathbf{Q}}_{2}\mathbf{\bar{\mathbf{H}}}_{1}^{H}(\mathbf{F}(\bar{\boldsymbol{\theta}}_{2})-\mathbf{F}(\bar{\boldsymbol{\theta}}_{1}))^{H}\mathbf{H}_{2}^{H}||\le ab\bar{P}_{t}||\bar{\boldsymbol{\theta}}_{1}-\bar{\boldsymbol{\theta}}_{2}||\label{eq:phi_second_part_1_1}
\end{multline}
and
\begin{gather}
||(\mathbf{Z}(\bar{\boldsymbol{\theta}}_{2})\bar{\mathbf{Q}}_{2}-\mathbf{Z}(\bar{\boldsymbol{\theta}}_{1})\bar{\mathbf{Q}}_{1})\mathbf{Z}(\bar{\boldsymbol{\theta}}_{1})^{H}||\nonumber \\
\le||\bar{\mathbf{H}}_{\mathrm{DIR}}(\bar{\mathbf{Q}}_{2}-\bar{\mathbf{Q}}_{1})\mathbf{Z}(\bar{\boldsymbol{\theta}}_{1})^{H}||\nonumber \\
+||\left[\mathbf{H}_{2}\mathbf{F}(\bar{\boldsymbol{\theta}}_{2})\bar{\mathbf{H}}_{1}\bar{\mathbf{Q}}_{2}-\mathbf{H}_{2}\mathbf{F}(\bar{\boldsymbol{\theta}}_{1})\bar{\mathbf{H}}_{1}\bar{\mathbf{Q}}_{1}\right]\mathbf{Z}(\bar{\boldsymbol{\theta}}_{1})^{H}||.\label{eq:phi_second_part_1_2}
\end{gather}
To upper-bound the two terms on the RHS of \eqref{eq:phi_second_part_1_2},
we use
\begin{equation}
\bigl\Vert\mathbf{\bar{\mathbf{H}}}_{\mathrm{DIR}}(\bar{\mathbf{Q}}_{2}-\bar{\mathbf{Q}}_{1})\mathbf{Z}(\bar{\boldsymbol{\theta}}_{1})^{H}\bigr\Vert\le b\lambda_{\max}(\bar{\mathbf{H}}_{\mathrm{DIR}})\bigl\Vert\bar{\mathbf{Q}}_{1}-\bar{\mathbf{Q}}_{2}\bigr\Vert\label{eq:phi_second_part_1_2_1}
\end{equation}
and
\begin{multline}
||\left[\mathbf{H}_{2}\mathbf{F}(\bar{\boldsymbol{\theta}}_{2})\mathbf{\bar{\mathbf{H}}}_{1}\bar{\mathbf{Q}}_{2}-\mathbf{H}_{2}\mathbf{F}(\bar{\boldsymbol{\theta}}_{1})\bar{\mathbf{H}}_{1}\bar{\mathbf{Q}}_{1}\right]\mathbf{Z}(\bar{\boldsymbol{\theta}}_{1})^{H}||\\
\le\lambda_{\max}(\mathbf{H}_{2})||\mathbf{F}(\bar{\boldsymbol{\theta}}_{2})\bar{\mathbf{H}}_{1}\bar{\mathbf{Q}}_{2}-\mathbf{F}(\bar{\boldsymbol{\theta}}_{1})\bar{\mathbf{H}}_{1}\bar{\mathbf{Q}}_{1}||\\
\times\left[\lambda_{\max}(\bar{\mathbf{H}}_{\mathrm{DIR}}^{H})+\lambda_{\max}(\mathbf{\bar{\mathbf{H}}}_{1}^{H})\lambda_{\max}(\mathbf{H}_{2}^{H})\right]\\
\le k^{-1}ab||\bar{\mathbf{Q}}_{1}-\bar{\mathbf{Q}}_{2}||+ab\bar{P}_{t}||\bar{\boldsymbol{\theta}}_{1}-\bar{\boldsymbol{\theta}}_{2}||.\label{eq:phi_second_part_1_2_2}
\end{multline}
Substituting \eqref{eq:phi_second_part_1_1}, \eqref{eq:phi_second_part_1_2},
\eqref{eq:phi_second_part_1_2_1} and \eqref{eq:phi_second_part_1_2_2}
into \eqref{eq:phi_second_part_1}, we obtain
\begin{multline}
||\mathbf{Z}(\bar{\boldsymbol{\theta}}_{2})\bar{\mathbf{Q}}_{2}\mathbf{Z}(\bar{\boldsymbol{\theta}}_{2})^{H}-\mathbf{Z}(\bar{\boldsymbol{\theta}}_{1})\bar{\mathbf{Q}}_{1}\mathbf{Z}(\bar{\boldsymbol{\theta}}_{1})^{H}||\\
\le b^{2}||\bar{\mathbf{Q}}_{1}-\bar{\mathbf{Q}}_{2}||+2ab\bar{P}_{t}||\bar{\boldsymbol{\theta}}_{1}-\bar{\boldsymbol{\theta}}_{2}||\label{eq:phi_second_part_1_fin}
\end{multline}
and \eqref{eq:phi_second_part} then implies
\begin{multline}
\!\!\!\!\!\!||\mathbf{H}_{2}^{H}\mathbf{K}(\bar{\boldsymbol{\theta}}_{1},\bar{\mathbf{Q}}_{1})\mathbf{Z}(\bar{\boldsymbol{\theta}}_{2})\bar{\mathbf{Q}}_{2}\bar{\mathbf{H}}_{1}^{H}-\mathbf{H}_{2}^{H}\mathbf{K}(\bar{\boldsymbol{\theta}}_{2},\bar{\mathbf{Q}}_{2})\mathbf{Z}(\bar{\boldsymbol{\theta}}_{2})\bar{\mathbf{Q}}_{2}\bar{\mathbf{H}}_{1}^{H}||\\
\le ab^{3}\bar{P}_{t}||\bar{\mathbf{Q}}_{1}-\bar{\mathbf{Q}}_{2}||+2a^{2}b^{2}\bar{P}_{t}^{2}||\bar{\boldsymbol{\theta}}_{1}-\bar{\boldsymbol{\theta}}_{2}||.\label{eq:phi_second_part_fin}
\end{multline}
Substituting \eqref{eq:phi_first_part_fin} and \eqref{eq:phi_second_part_fin}
into \eqref{eq:grad_phi_dif}, we obtain \eqref{eq:grad_phi_dif_fin}.}

\subsection{\textcolor{black}{Proof of \eqref{eq:grad_Q_dif_fin}}}

\textcolor{black}{From \eqref{eq:grad_Q} immediately have
\begin{gather}
||\nabla_{\bar{\mathbf{Q}}}f(\bar{\boldsymbol{\theta}}_{1},\bar{\mathbf{Q}}_{1})-\nabla_{\bar{\mathbf{Q}}}f(\bar{\boldsymbol{\theta}}_{2},\bar{\mathbf{Q}}_{2})||\nonumber \\
=||\mathbf{Z}(\bar{\boldsymbol{\theta}}_{1})^{H}\mathbf{K}(\bar{\boldsymbol{\theta}}_{1},\bar{\mathbf{Q}}_{1})\mathbf{Z}(\bar{\boldsymbol{\theta}}_{1})-\mathbf{Z}(\bar{\boldsymbol{\theta}}_{2})^{H}\mathbf{K}(\bar{\boldsymbol{\theta}}_{2},\bar{\mathbf{Q}}_{2})\mathbf{Z}(\bar{\boldsymbol{\theta}}_{2})||\nonumber \\
\le||\mathbf{Z}(\bar{\boldsymbol{\theta}}_{1})^{H}\mathbf{K}(\bar{\boldsymbol{\theta}}_{1},\bar{\mathbf{Q}}_{1})\mathbf{Z}(\bar{\boldsymbol{\theta}}_{1})-\mathbf{Z}(\bar{\boldsymbol{\theta}}_{1})^{H}\mathbf{K}(\bar{\boldsymbol{\theta}}_{1},\bar{\mathbf{Q}}_{1})\mathbf{Z}(\bar{\boldsymbol{\theta}}_{2})||\nonumber \\
+||\mathbf{Z}(\bar{\boldsymbol{\theta}}_{1})^{H}\mathbf{K}(\bar{\boldsymbol{\theta}}_{1},\bar{\mathbf{Q}}_{1})\mathbf{Z}(\bar{\boldsymbol{\theta}}_{2})-\mathbf{Z}(\bar{\boldsymbol{\theta}}_{2})^{H}\mathbf{K}(\bar{\boldsymbol{\theta}}_{2},\bar{\mathbf{Q}}_{2})\mathbf{Z}(\bar{\boldsymbol{\theta}}_{2})||.\label{eq:grad_Q_dif}
\end{gather}
}%
\begin{comment}
\textcolor{black}{At this point the reviewer will know that we will
follow the same steps to prove \eqref{eq:grad_phi_dif_fin1}, so we
simply say that and skip the rest. If asked, we provide it.}
\end{comment}
\textcolor{black}{Following the same steps used to prove \eqref{eq:grad_phi_dif_fin}
we can further upper bound the two norms in the RHS of the above equation
to prove \eqref{eq:grad_Q_dif_fin}. The details are omitted here
due to the page limit.}

\section{\textcolor{black}{Proof of Theorem \ref{thm:convergence} \label{sec:convergence}}}

\textcolor{black}{We recall the following inequality for any function
$f(x)$ which is $L$-smooth:
\begin{equation}
f(\mathbf{y})\geq f(\mathbf{x})+\bigl\langle\nabla f\bigl(\mathbf{x}\bigr),\mathbf{y}-\mathbf{x}\bigr\rangle-\frac{L}{2}||\mathbf{y}-\mathbf{x}||^{2}.\label{eq:Lips:grad:inequality}
\end{equation}
The projection of $\bar{\boldsymbol{\theta}}_{n+1}$ onto $\bar{\Theta}$
can be written as
\begin{multline}
\bar{\boldsymbol{\theta}}_{n+1}=\underset{\bar{\boldsymbol{\theta}}\in\bar{\Theta}}{\arg\min}\bigl\Vert\bar{\boldsymbol{\theta}}-\bar{\boldsymbol{\theta}}_{n}-\mu\nabla_{\bar{\boldsymbol{\theta}}}f\bigl(\bar{\boldsymbol{\theta}}_{n},\bar{\mathbf{Q}}_{n}\bigr)\bigr)\bigr\Vert^{2}\\
=\underset{\bar{\boldsymbol{\theta}}\in\bar{\Theta}}{\arg\max}\ \bigl\langle\nabla_{\bar{\boldsymbol{\theta}}}f\bigl(\bar{\boldsymbol{\theta}}_{n},\bar{\mathbf{Q}}_{n}\bigr),\bar{\boldsymbol{\theta}}-\bar{\boldsymbol{\theta}}_{n}\bigr\rangle-\frac{1}{2\mu}||\bar{\boldsymbol{\theta}}-\bar{\boldsymbol{\theta}}_{n}||^{2}\label{eq:theta:project:rewrite}
\end{multline}
where $\bigl\langle\mathbf{x},\mathbf{y}\bigr\rangle=\Re(\mathbf{x}^{H}\mathbf{y})$
and we have used the fact that $||\mathbf{a}-\mathbf{b}||^{2}=||\mathbf{a}||^{2}+||\mathbf{b}||^{2}-2\Re(\mathbf{a}^{H}\mathbf{b})$.
Note that when $\bar{\boldsymbol{\theta}}=\bar{\boldsymbol{\theta}}_{n}$,
the objective in the above problem is equal to 0, and thus we have
\begin{equation}
\bigl\langle\nabla_{\bar{\boldsymbol{\theta}}}f\bigl(\bar{\boldsymbol{\theta}}_{n},\bar{\mathbf{Q}}_{n}\bigr),\bar{\boldsymbol{\theta}}_{n+1}-\bar{\boldsymbol{\theta}}_{n}\bigr\rangle-\frac{1}{2\mu}||\bar{\boldsymbol{\theta}}_{n+1}-\bar{\boldsymbol{\theta}}_{n}||^{2}\geq0.
\end{equation}
An analogous inequality also holds for $\bar{\mathbf{Q}}_{n+1}$,
i.e.,
\begin{equation}
\bigl\langle\nabla_{\bar{\mathbf{Q}}}f\bigl(\bar{\boldsymbol{\theta}}_{n},\bar{\mathbf{Q}}_{n}\bigr),\bar{\mathbf{Q}}_{n+1}-\bar{\mathbf{Q}}_{n}\bigr\rangle-\frac{1}{2\mu}||\bar{\mathbf{Q}}_{n+1}-\bar{\mathbf{Q}}_{n}||^{2}\geq0.
\end{equation}
Applying \eqref{eq:Lips:grad:inequality} yields
\begin{align}
f(\bar{\boldsymbol{\theta}}_{n+1},\bar{\mathbf{Q}}_{n+1}) & \geq f\bigl(\bar{\boldsymbol{\theta}}_{n},\bar{\mathbf{Q}}_{n}\bigr)+\bigl\langle\nabla_{\bar{\boldsymbol{\theta}}}f\bigl(\bar{\boldsymbol{\theta}}_{n},\bar{\mathbf{Q}}_{n}\bigr),\bar{\boldsymbol{\theta}}_{n+1}-\bar{\boldsymbol{\theta}}_{n}\bigr\rangle\nonumber \\
 & +\bigl\langle\nabla_{\bar{\mathbf{Q}}}f\bigl(\bar{\boldsymbol{\theta}}_{n},\bar{\mathbf{Q}}_{n}\bigr),\bar{\mathbf{Q}}_{n+1}-\bar{\mathbf{Q}}_{n}\bigr)\bigr\rangle\nonumber \\
 & -\frac{L}{2}\bigl\Vert\bar{\boldsymbol{\theta}}_{n+1}-\bar{\boldsymbol{\theta}}_{n}\bigr\Vert^{2}-\frac{L}{2}\bigl\Vert\bar{\mathbf{Q}}_{n+1}-\bar{\mathbf{Q}}_{n}\bigr\Vert^{2}\nonumber \\
 & \geq f\bigl(\bar{\boldsymbol{\theta}}_{n},\bar{\mathbf{Q}}_{n}\bigr)+\bigl(\frac{1}{2\mu}-\frac{L}{2}\bigr)\bigl(\bigl\Vert\bar{\boldsymbol{\theta}}_{n+1}-\bar{\boldsymbol{\theta}}_{n}\bigr\Vert^{2}\nonumber \\
 & +\bigl\Vert\bar{\mathbf{Q}}_{n+1}-\bar{\mathbf{Q}}_{n}\bigr\Vert^{2}\bigr).\label{eq:theta:project:ineq}
\end{align}
It is easy to see that $f(\bar{\boldsymbol{\theta}}_{n+1},\bar{\mathbf{Q}}_{n+1})\geq f\bigl(\bar{\boldsymbol{\theta}}_{n},\bar{\mathbf{Q}}_{n}\bigr)$
if $\mu<\frac{1}{L}$. Since the feasible set of the considered problem
is closed and bounded, the iterate $(\bar{\boldsymbol{\theta}}_{n},\bar{\mathbf{Q}}_{n})$
is bounded and thus $\bigl(\bar{\boldsymbol{\theta}}_{n},\bar{\mathbf{Q}}_{n}\bigr)$
has accumulation points. Since, as shown above, $f\bigl(\bar{\boldsymbol{\theta}}_{n},\bar{\mathbf{Q}}_{n}\bigr)$
is nondecreasing, $f$ has the same value, denoted by $f^{\ast}$,
at all of these accumulation points. From \eqref{eq:theta:project:ineq}
we have
\begin{multline}
f\bigl(\bar{\boldsymbol{\theta}}_{n+1},\bar{\mathbf{Q}}_{n+1}\bigr)-f\bigl(\bar{\boldsymbol{\theta}}_{n},\bar{\mathbf{Q}}_{n}\bigr)\geq\bigl(\frac{1}{2\mu}-\frac{L}{2}\bigr)\bigl(\bigl\Vert\bar{\boldsymbol{\theta}}_{n+1}-\bar{\boldsymbol{\theta}}_{n}\bigr\Vert^{2}\\
+\bigl\Vert\bar{\mathbf{Q}}_{n+1}-\bar{\mathbf{Q}}_{n}\bigr\Vert^{2}\bigr),
\end{multline}
which results in
\begin{multline}
\infty>f^{\ast}-f\bigl(\bar{\boldsymbol{\theta}}_{1},\bar{\mathbf{Q}}_{1}\bigr)\geq\sum_{n=1}^{\infty}\bigl(\frac{1}{2\mu}-\frac{L}{2}\bigr)\bigl(\bigl\Vert\bar{\boldsymbol{\theta}}_{n+1}-\bar{\boldsymbol{\theta}}_{n}\bigr\Vert^{2}\\
+\bigl\Vert\bar{\mathbf{Q}}_{n+1}-\bar{\mathbf{Q}}_{n}\bigr\Vert^{2}\bigr).
\end{multline}
Since $\mu<\frac{1}{L}$ we can conclude that
\begin{equation}
\bigl\Vert\bar{\boldsymbol{\theta}}_{n+1}-\bar{\boldsymbol{\theta}}_{n}\bigr\Vert\to0;\bigl\Vert\bar{\mathbf{Q}}_{n+1}-\bar{\mathbf{Q}}_{n}\bigr\Vert\to0.\label{eq:iterate:converge}
\end{equation}
The optimality condition of \eqref{eq:theta:project:rewrite} implies
\begin{equation}
\bigl\langle\frac{1}{\mu}\bigl(\bar{\boldsymbol{\theta}}_{n+1}-\bar{\boldsymbol{\theta}}_{n}\bigr)-\nabla_{\bar{\boldsymbol{\theta}}}f\bigl(\bar{\boldsymbol{\theta}}_{n},\bar{\mathbf{Q}}_{n}\bigr),\bar{\boldsymbol{\theta}}-\bar{\boldsymbol{\theta}}_{n+1}\bigr\rangle\leq0,\ \forall\bar{\boldsymbol{\theta}}\in\bar{\Theta}.\label{eq:optimalaity:theta}
\end{equation}
Similarly we have
\begin{equation}
\bigl\langle\frac{1}{\mu}\bigl(\bar{\mathbf{Q}}_{n+1}-\bar{\mathbf{Q}}_{n}\bigr)-\nabla_{\bar{\mathbf{Q}}}f\bigl(\bar{\boldsymbol{\theta}}_{n},\bar{\mathbf{Q}}_{n}\bigr),\bar{\mathbf{Q}}-\bar{\mathbf{Q}}_{n+1}\bigr\rangle\leq0,\ \forall\bar{\mathbf{Q}}\in\bar{\mathcal{Q}}.\label{eq:optimiality:Q}
\end{equation}
Let $\bigl(\boldsymbol{\theta}^{\ast},\mathbf{Q}^{\ast}\bigr)$ be
any accumulation point of $\bigl(\bar{\boldsymbol{\theta}}_{n},\bar{\mathbf{Q}}_{n}\bigr)$,
say $\bigl(\bar{\boldsymbol{\theta}}_{n},\bar{\mathbf{Q}}_{n}\bigr)\to\bigl(\boldsymbol{\theta}^{\ast},\mathbf{Q}^{\ast}\bigr)$
as $n\to\infty$. We also note that the gradient of $f\bigl(\bar{\boldsymbol{\theta}}_{n},\bar{\mathbf{Q}}_{n}\bigr)$
is continuous and thus $\nabla_{\bar{\boldsymbol{\theta}}}f\bigl(\bar{\boldsymbol{\theta}}_{n},\bar{\mathbf{Q}}_{n}\bigr)\to\nabla_{\bar{\boldsymbol{\theta}}}f\bigl(\boldsymbol{\theta}^{\ast},\mathbf{Q}^{\ast}\bigr)$
and $\nabla_{\bar{\mathbf{Q}}}f\bigl(\bar{\boldsymbol{\theta}}_{n},\bar{\mathbf{Q}}_{n}\bigr)\to\nabla_{\bar{\mathbf{Q}}}f\bigl(\boldsymbol{\theta}^{\ast},\mathbf{Q}^{\ast}\bigr)$.
By letting $n\to\infty$ in \eqref{eq:optimalaity:theta} and \eqref{eq:optimiality:Q},
we have
\begin{equation}
\bigl\langle-\nabla_{\bar{\boldsymbol{\theta}}}f\bigl(\boldsymbol{\theta}^{\ast},\mathbf{Q}^{\ast}\bigr),\bar{\boldsymbol{\theta}}-\boldsymbol{\theta}^{\ast}\bigr\rangle\leq0,\ \forall\bar{\boldsymbol{\theta}}\in\bar{\Theta}
\end{equation}
\begin{equation}
\bigl\langle-\nabla_{\bar{\mathbf{Q}}}f\bigl(\boldsymbol{\theta}^{\ast},\mathbf{Q}^{\ast}\bigr),\bar{\mathbf{Q}}-\mathbf{Q}^{\ast}\bigr\rangle\leq0,\ \forall\bar{\mathbf{Q}}\in\bar{\mathcal{Q}},
\end{equation}
which means that $\bigl(\boldsymbol{\theta}^{\ast},\mathbf{Q}^{\ast}\bigr)$
is indeed a critical point of \eqref{eq:capacityprob}. This completes
the proof.}

\section{Computational Complexity for Alternating Optimization (AO) \label{sec:AO_comp}}

The computational complexity for the AO method, introduced in \cite{zhang2019capacity},
is derived in this appendix. To make the following derivation more
accessible, the mathematical notation in this appendix is the same
as in \cite{zhang2019capacity}.

The channel matrix from the transmitter to the receiver is given by
$\tilde{\mathbf{H}}=\mathbf{\mathbf{H}}+\mathbf{R}\mathbf{\boldsymbol{\phi}T}$,
where $\mathbf{H}\in\mathbb{C}^{N_{r}\times N_{t}}$ presents the
\emph{direct} signal transmission between the transmitter and the
receiver, $\mathbf{T}\in\mathbb{C}^{N_{\mathrm{ris}}\times N_{t}}$
presents the signal transmission between the transmitter and the RIS,
$\mathbf{R}\in\mathbb{C}^{N_{r}\times N_{\mathrm{ris}}}$ presents
the signal transmission between the RIS and the receiver, and $\boldsymbol{\phi}$
models the RIS response. Let $\mathbf{\mathbf{R}}=[\mathbf{r}_{1},\dots,\mathbf{r}_{N_{\mathrm{ris}}}]$,
$\mathbf{\mathbf{T}}=[\mathbf{t}_{1},\dots,\mathbf{t}_{N_{\mathrm{ris}}}]^{H}$
and $\mathbf{\mathbf{\boldsymbol{\phi}}}=\diag[\alpha_{1},\dots,\alpha_{N_{\mathrm{ris}}}]$,
so that the channel matrix can be written as $\tilde{\mathbf{H}}=\mathbf{\mathbf{H}}+\sum_{i=1}^{N_{\mathrm{ris}}}\alpha_{i}\mathbf{r}_{i}\mathbf{t}_{i}^{H}$.

In the first step of the AO algorithm, $L_{\mathrm{AO}}$ independent
realizations of $\{\alpha_{m}\}_{m=1}^{N_{\mathrm{ris}}}$ are randomly
generated and for each of these the optimal covariance matrix $\mathbf{Q}$
is computed. To do this, the channel matrix $\tilde{\mathbf{H}}$
has to be calculated for every $\{\alpha_{m}\}_{m=1}^{N_{\mathrm{ris}}}$
realization. This calculation starts by computing all $\mathbf{r}_{m}\mathbf{t}_{m}^{H}$
matrices and for this $N_{r}N_{t}N_{\mathrm{ris}}$ multiplications
are needed. Further, the computation of all $\alpha_{m}\mathbf{r}_{m}\mathbf{t}_{m}^{H}$
matrices requires $N_{r}N_{t}N_{\mathrm{ris}}$ multiplications (per
one $\{\alpha_{m}\}_{m=1}^{N_{\mathrm{ris}}}$ realization). Hence,
the complexity of calculating $L_{\mathrm{AO}}$ channel matrices
$\tilde{\mathbf{H}}$ is $(L_{\mathrm{AO}}+1)N_{r}N_{t}N_{\mathrm{ris}}$.

For each $\tilde{\mathbf{H}}$ it is required to perform the the truncated
singular value decomposition $\tilde{\mathbf{H}}=\tilde{\mathbf{U}}\tilde{\boldsymbol{\varLambda}}\tilde{\mathbf{V}}^{H}$,
where $\tilde{\mathbf{V}}\in\mathbb{C}^{N_{t}\times D}$ and $D=\min(N_{t},N_{r})$.
The complexity of this decomposition is approximately $\mathcal{O}(D^{3})$.
Next, $\mathbf{Q}$ is computed as $\mathbf{Q}=\tilde{\mathbf{V}}\diag\{p_{1},\dots,p_{D}\}\tilde{\mathbf{V}}^{H}$,
where $\{p_{1},\dots,p_{D}\}$ are obtained using a water-filling
algorithm and $D=\min(N_{t},N_{r})$. The complexity of the water-filling
algorithm is $\mathcal{O}(D^{2})$ and the complexity of the matrix
multiplication is $N_{t}D+(N_{t}^{2}+N_{t})D/2$. Therefore, the calculation
of $L_{\mathrm{AO}}$ covariance matrices $\mathbf{Q}$ requires $\mathcal{O}(L_{\mathrm{AO}}(D^{3}+\frac{1}{2}N_{t}^{2}D))$
multiplications.

In the sequel, the optimal $\{\alpha_{m}\}_{m=1}^{N_{\mathrm{ris}}}$
and $\mathbf{Q}$ are iteratively determined. In one conventional
iteration, one $\alpha_{m}$ or $\mathbf{Q}$ is adjusted. A set of
$N_{\mathrm{ris}}+1$ successive conventional iterations constitutes
one ``outer'' iteration, in which \emph{all} $\alpha_{m}$ and $\mathbf{Q}$
are adjusted. The AO method stops when the convergence criterion at
the end of an outer iteration is fulfilled.

At the beginning of each outer iteration, the eigenvalue decomposition
$\mathbf{Q}=\mathbf{U}_{Q}\boldsymbol{\Sigma}_{Q}\mathbf{U}_{Q}^{H}$
is performed, which requires $\mathcal{O}(N_{t}^{3})$ multiplications.
The calculation of the matrices $\mathbf{H}'=\mathbf{H}\mathbf{U}_{Q}\boldsymbol{\Sigma}_{Q}^{\frac{1}{2}}\in\mathbb{C}^{N_{r}\times N_{t}}$
and $\mathbf{\mathbf{T}}'=\mathbf{T}\mathbf{U}_{Q}\boldsymbol{\Sigma}_{Q}^{\frac{1}{2}}\in\mathbb{C}^{N_{\mathrm{ris}}\times N_{t}}$
has the complexity $\mathcal{O}(N_{t}^{3}+N_{t}^{2}N_{\mathrm{ris}})$.

To form $\mathbf{S}=\mathbf{H}'+\sum_{i=1}^{N_{\mathrm{ris}}}\alpha_{i}\mathbf{r}_{i}\mathbf{t}_{i}^{'H}$,
$N_{r}N_{t}N_{\mathrm{ris}}$ multiplications are needed first to
obtain all $\mathbf{r}_{i}\mathbf{t}_{i}^{'H}$ and $N_{r}N_{t}N_{\mathrm{ris}}$
multiplications are needed for all $\alpha_{i}\mathbf{r}_{i}\mathbf{t}_{i}^{'H}$.
Hence, the complexity of computing $\mathbf{S}$ is $2N_{r}N_{t}N_{\mathrm{ris}}$.

The optimization of the $m$-th RIS element requires the computation
of the following auxiliary matrices
\begin{align}
\mathbf{A}_{m} & =\mathbf{I}+\frac{1}{N_{0}}\mathbf{S}_{m}\mathbf{S}_{m}^{H}+\frac{1}{N_{0}}\mathbf{r}_{m}\mathbf{t}_{m}^{'H}\left(\mathbf{r}_{m}\mathbf{t}_{m}^{'H}\right)^{H}\\
\mathbf{B}_{m} & =\frac{1}{N_{0}}\mathbf{r}_{m}\mathbf{t}_{m}^{'H}\mathbf{S}_{m}^{H}
\end{align}
where $\mathbf{S}_{m}=\mathbf{H}'+\sum_{i=1,i\neq m}^{N_{\mathrm{ris}}}\alpha_{i}\mathbf{r}_{i}\mathbf{t}_{i}^{'H}=\mathbf{S}-\alpha_{m}\mathbf{r}_{m}\mathbf{t}_{m}^{'H}$.
It can be easily shown that the complexities for calculating $\mathbf{A}_{m}$
and $\mathbf{B}_{m}$ from the previous expressions are both $\mathcal{O}(N_{r}^{2}N_{t})$.

Utilizing the results from Subsection \ref{subsec:PGM_comp}, the
complexity of computing $\mathbf{A}_{m}^{-1}\mathbf{B}_{m}$ is $\mathcal{O}(2N_{r}^{3})$.
The subsequent calculation of $\alpha_{m}$ and update of $\mathbf{S}=\mathbf{S}_{m}+\alpha_{m}\mathbf{r}_{m}\mathbf{t}_{m}^{'H}$
require a negligible complexity.

After adjusting all RIS elements, the optimization of $\mathbf{\mathbf{Q}}$
is performed according to the aforementioned procedure, which requires
$\mathcal{O}(D^{3}+\frac{1}{2}N_{t}^{2}D)$ multiplications.

If $I_{\mathrm{OI}}$ is the number of outer iterations, then the
computational complexity of the AO algorithm is given by
\begin{gather}
C_{\mathrm{AO}}=\mathcal{O}((L_{\mathrm{AO}}+1)N_{r}N_{t}N_{\mathrm{ris}}+L_{\mathrm{AO}}(D^{3}+\frac{1}{2}N_{t}^{2}D)\nonumber \\
+I_{\mathrm{OI}}[N_{t}^{3}+N_{t}^{2}N_{\mathrm{ris}}+2N_{r}N_{t}N_{\mathrm{ris}}\nonumber \\
+(2N_{r}^{2}N_{t}+2N_{r}^{3})N_{\mathrm{ris}}+D^{3}+\frac{1}{2}N_{t}^{2}D])\label{eq:ao_comp1}
\end{gather}
where $D=\min(N_{t},N_{r})$.

\textcolor{black}{}

\textcolor{black}{\bibliographystyle{IEEEtran}
\bibliography{IEEEabrv,IEEEexample,RIS_rate_fin_Arxiv}
}
\end{document}